\documentclass[aps, twocolumn, nofootinbib, showpacs, prd]{revtex4}
\usepackage{graphicx}
\usepackage{amsfonts}
\usepackage{amssymb}
\usepackage{amsbsy}
\usepackage{amsmath}
\usepackage{mathtools}
\usepackage{latexsym}
\usepackage{natbib}
\usepackage{bm}
\usepackage{ulem}
\usepackage[utf8]{inputenc}
\usepackage{newunicodechar}
\newunicodechar{∼}{$\sim$}
\usepackage{color}
\usepackage{float}
\usepackage{hyperref}
\usepackage{multirow}
\usepackage{array,makecell}

\begin{document}
\title{Chiellini-Integrable Cosmologies with Phantom Divide Crossing}

\author{Soumya Chakrabarti}
\email{soumya.chakrabarti@vit.ac.in}
\affiliation{School of Advanced Sciences, Vellore Institute of Technology, \\ 
Tiruvalam Rd, Katpadi, Vellore, Tamil Nadu 632014 \\ India}

\author{Nandan Roy}
\email{nandan.roy@mahidol.edu}
\affiliation{Centre for Theoretical Physics and Natural Philosophy, Nakhonsawan Studiorum for Advanced Studies, Mahidol University, Thailand}


\date{\today}

\begin{abstract}
We investigate exact cosmological solutions of a massive scalar field minimally coupled to General Relativity. For an extended Higgs-like scalar self-interaction, we find that the resulting field equations belong to the damped Ermakov-Painlev\'e II class and construct novel analytical solutions within the framework of the Chiellini integrability condition. We analyze the expanding branch of the solutions in the context of late-time cosmic acceleration, using a combined statistical analysis of BAO, CMB, cosmic chronometer and Pantheon+SHOES supernova datasets. A crucial outcome of this exercise is an analytical emergence of the smooth crossing of phantom divide in the dark energy equation of state, achieved without introducing any pathological instabilities. The reconstruction yields a present-day Hubble parameter $H_0 \gtrsim 70 \,\mathrm{km\,s^{-1}\,Mpc^{-1}}$, with a reduced tension relative to the $\Lambda$CDM cosmology. The results indicate that Chiellini-integrable scalar cosmologies are capable of providing a robust and analytically controlled framework for modeling late-time cosmic acceleration and phantom divide crossing, offering a viable alternative to phenomenological dark-energy parametrizations.
\end{abstract}

\maketitle

\section{Introduction}\label{sec:introduction}
The study of non-linear differential equations has long played a central role in mathematical physics. It forms a crucial field of research, providing insight into complex dynamical systems that can not be adequately described by linear models. Among these, a particularly important family is represented by the Milne-Pinney class of equations \cite{Milne, Pinney}, which has its conceptual origins in the pioneering work of Ermakov \cite{Ermakov}. Over the past century, the Milne-Ermakov-Pinney (MEP) equation has been reformulated in diverse contexts, appearing in areas ranging from nonlinear optics, plasma physics to quantum mechanics \cite{bini2020new, esposito2019new, Haas, Carinena, haas2021relativistic, cariglia2018cosmological, herring2007feshbach, PhysRevD.105.106023}. The canonical form of the MEP equation is given by
\begin{equation}
\ddot{y} + \omega^2(t) y = \frac{\kappa}{y^3}, \label{eq:MEP}
\end{equation}
where $\omega(t)$ is a time-dependent frequency function and $\kappa$ is a real constant related to the invariant structure of the system. This differential equation may be regarded as a non-linear deformation of the linear oscillator equation $\ddot{x} + \omega^2(t)x = 0$ and a general solution can be constructed based on the solutions of the associated linear oscillator \cite{mukherjee2016generalized, carinena2009applications, haas2021relativistic}. Taking into account dissipative or damping effects that are often physically relevant one can write the damped Milne-Ermakov-Pinney equation as
\begin{equation}
\ddot{y} + \mu \dot{y} + \omega^2(t)y = \frac{\kappa}{y^3}, \qquad \mu > 0, \label{eq:DMEP}
\end{equation}
where $\mu$ represents a positive damping coefficient. Under suitable transformations, Eq. \eqref{eq:DMEP} can be mapped into a generalized Emden-Fowler equation of index $-3$, a form known to admit exact integrability under certain constraints \cite{coffman, MELLIN1994529, BOHMER01012010}. Nonlinear equations of this kind naturally connect to other important families of integrable systems, for instance, a hybrid Ermakov-Painlev\'e~II system is obtained as a reduction of the coupled $(N+1)$-dimensional nonlinear Schr\"odinger system \cite{Rogers}. This hierarchy occupies a central place in modern mathematical physics due to applications in random matrix theory, soliton dynamics, and nonlinear wave propagation \cite{carillo, Morris_2015, PANAYOTOUNAKOS2006634, CHOWDHURY2009104}. Parallel to these developments, renewed attention has been directed towards a Chiellini-type integrability condition \cite{AC}, which can establish a specific functional relationship between the damping and potential term in a damped Milne-Ermakov-Pinney equation and guarantees exact integrability \cite{AL}. Within the broader context of Hamiltonization techniques, this technique naturally arises through a Jacobi last-multiplier method \cite{Nucci_2008, GHOSECHOUDHURY2009651, sym15071416}. Works by Bandic \cite{IB1, IB2}, Mak, Harko, and their collaborators \cite{MH1, MH2, MH3, YY, MH4}, as well as by Mancas and Rosu \cite{RMC, MR, MR1}, have systematically extended the applicability of this condition, revealing new classes of solvable nonlinear systems in fluid mechanics and quantum dissipative models.  \\

We investigate the applicability of a Chiellini integrability condition in General Relativity (GR), in order to construct cosmological models for the present universe. It is well known that the gravitational field equations in a variety of settings can be recast into the form of damped Ermakov-Milne-Pinney equations and related hybrid forms \cite{Hawkins_2002, GhoseChoudhury, PhysRevD.95.024015, Batic_2023}. In this context, the integrability of first-order Abel equations \cite{AD}, second-order Lienard-type equations \cite{AL} and the associated techniques can also play a central role \cite{HT, CM}. Building upon this correspondence, we demonstrate that the Klein-Gordon equation for a spatially homogeneous scalar field minimally coupled to GR can be reformulated exactly as a damped Ermakov-Painlev\'e~II equation, admitting analytical solutions within a unified Chiellini framework. We show that the resulting exact solution naturally gives rise to an accelerated expansion of the universe. Exploiting the analytic invertibility of the solution, we derive an effective Hubble function that can be expressed as a systematic extension of the standard $\Lambda$CDM cosmology. We confront this framework with a combined statistical analysis of the BAO, CMB, cosmic chronometer (CC) and Pantheon+SHOES datasets. The analysis is supplemented by a second-order Pad\'e approximation and yields a present-day Hubble parameter $H_0 \gtrsim 70\,\mathrm{km\,s^{-1}\,Mpc^{-1}}$, with a reduced tension relative to the $\Lambda$CDM model while remaining consistent with current cosmological constraints. We also find a smooth and controlled crossing of the $w = -1$ barrier for the dark energy equation of state, as a direct consequence of the underlying integrable dynamics, thereby providing a mathematically grounded realization of the so-called late-time phantom crossing. \\

It is important to emphasize that the Chiellini integrability condition is not a generic solvability criterion, but a restrictive structural constraint that singles out a distinguished and nontrivial sub-class of nonlinear dissipative systems. Unlike phenomenological reconstructions, which typically rely on adjustable functional forms and truncated expansions, the Chiellini condition enforces an exact functional relation between the damping and restoring terms of the dynamical equation. As a consequence, integrability is achieved without sacrificing non-linearity or dissipation, and the resulting solutions are obtained in closed analytical form. This rigidity renders the framework exceptionally robust : once the Chiellini condition is satisfied, the qualitative behavior of the system : such as the existence of accelerated expansion, is protected against small deformations of initial conditions or parametrization choices. In a cosmological context, this places Chiellini-integrable models in sharp contrast with conventional dark-energy parametrizations, providing a rare example where late-time cosmic acceleration and its observationally inferred properties emerge from an underlying exact and structurally stable nonlinear dynamics rather than from phenomenological tuning.

\section{Cosmology from a Chiellini-Integrable Damped Nonlinear System}

\medskip
\noindent\textbf{Integrability 1.}\\
(a) If the damped Ermakov-Painlev\'e~II equation
\begin{equation}
\ddot{y} + g(y)\dot{y} + h(y) = 0 , \qquad h(y) = \lambda y + \epsilon y^3 - \frac{\eta}{y^3}, \label{eq:DEP}
\end{equation}
satisfies the Chiellini condition,
\begin{equation}
\frac{d}{dy}\!\left\lbrace \frac{h(y)}{g(y)}\right\rbrace = R_{0}\, g(y), \qquad R_{0} \in \mathbb{R},
\end{equation}
an analytic solvability is ensured. Explicit solutions include
\begin{equation}
y(t) = \sqrt{\frac{1}{(t-t_0)^2} + \sqrt{\tfrac{p}{3}}}, \qquad \epsilon = -1,
\end{equation}
and
\begin{equation}
y(t) = \left(\frac{p^2}{16\lambda^2} - \frac{\eta}{\lambda}\right)^{\!1/2} \sin\!\Big[2\sqrt{2\lambda}(t-t_0)\Big] + \frac{p}{4\lambda}, \qquad \epsilon = 0.
\end{equation}

Eq. \eqref{eq:DEP} combines three characteristic nonlinearities : the Painlev\'e-type cubic term $\epsilon y^3$, a linear restoring term $\lambda y$, and an inverse-cubic Ermakov contribution $-\eta/y^3$. The Chiellini condition enforces a functional relation between the damping $g(y)$ and $h(y)$, reducing the system to a first-order separable form. This balance between dissipation and nonlinearity generalizes constant-damping scenarios and guarantees integrability. For $\epsilon=-1$, the solution exhibits an inverse-square-root decay; for $\epsilon = 0$, it yields oscillatory motion with effective frequency $\sqrt{2\lambda}$. The singular term $\eta/y^3$ modifies amplitude and phase, introducing nonlinear frequency shifts. These parameter-dependent regimes highlight how the same dynamical form can generate distinct physical behaviors.

\medskip
\noindent\textbf{Integrability 2.} \\
(b) The generalized damped Milne-Pinney equation
\begin{equation}
\ddot{y} + g(y)\dot{y} + \lambda y = \frac{k_1}{y^3} + \frac{k_2}{y^2} + \sum_{n=0}^{R} \delta_n y^{2n+1},\label{eq:GDMP}
\end{equation}
under the Chiellini constraint, admits for $R = 0$ the parametric solution
\begin{align}
    t &= y_0 \omega 
    + \frac{f'(y_0)}{4\wp'(\omega_0)}
    \Bigg[\ln\!\frac{\sigma(\omega+\hat{c}-\omega_0)}{\sigma(\omega+\hat{c}+\omega_0)} 
    + 2(\omega+\hat{c})\zeta(\omega_0)\Bigg] 
    + \delta, \nonumber \\
    y &= y_0 + \frac{f'(y_0)}{4\big[\wp(\omega+\hat{c}) - f''(y_0)/24\big]},
\end{align}
where $f(y) = 2(\delta_0 - \lambda)y^4 + cy^2 - 4k_2y - 2k_1$, $y_0$ is a root of $f(y)=0$ and $\wp, \sigma, \zeta$ denote Weierstrass elliptic functions. The appearance of elliptic functions underscores the algebraic depth of the GDMP system. The invariants of the cubic polynomial $f(y)$ govern the amplitude and periodicity of the motion, with $f(y_0)=0$ fixing the oscillation scale and $\wp(\omega)$ encoding its phase structure. Such elliptic behavior naturally arises in nonlinear oscillators containing both polynomial and inverse-power terms, linking Ermakov-type dynamics with classical elliptic potentials.  \\

Overall, Eqs. \eqref{eq:DEP} and \eqref{eq:GDMP} illustrate the usefulness of the Chiellini approach : by imposing a specific differential relation between damping and potential terms, one can derive closed-form analytic solutions even for nonlinear dissipative systems. We explore the utility of \textit{Integrability 1} in solving the field equations of a spatially homogeneous scalar field minimally coupled to gravity through the Einstein-Hilbert action,
\begin{equation}\label{action1}
\mathcal{A}=\int d^4x \sqrt{-g}\left[R + \frac{1}{2} g^{\mu\nu}\partial_{\mu}\phi \partial_{\nu}\phi - V(\phi) + \mathcal{L}_{m}\right],
\end{equation}
where $R$ denotes the Ricci scalar, $\phi$ is a canonical scalar field with self-interaction potential $V(\phi)$, and $\mathcal{L}_{m}$ represents the Lagrangian density of ordinary matter. Varying the action with respect to the metric tensor yields the energy-momentum tensor associated with the scalar field,
\begin{equation}\label{minimallyscalar}
T^{\phi}_{\mu\nu} = \partial_{\mu}\phi \partial_{\nu}\phi - g_{\mu\nu} \left[\frac{1}{2} g^{\alpha\beta}\partial_{\alpha}\phi \partial_{\beta}\phi - V(\phi) \right].
\end{equation}

For a homogeneous and isotropic Universe described by the spatially flat Friedmann-Lemaitre-Robertson-Walker (FLRW) metric,
\begin{equation}
ds^{2} = -dt^{2} + a^{2}(t)\left(dx^{2}+dy^{2}+dz^{2}\right),
\end{equation}
with scale factor $a(t)$, the Einstein field equations can be derived as
\begin{equation}\label{fe1minimal}
3\left(\frac{\dot{a}}{a}\right)^{2} = \rho_{m} + \rho_{\phi} = \rho_{m} + \frac{\dot{\phi}^{2}}{2} + V(\phi),
\end{equation}
and
\begin{equation}\label{fe2minimal}
-2\frac{\ddot{a}}{a} - \left(\frac{\dot{a}}{a}\right)^{2} = p_{m} + p_{\phi} = p_{m} + \frac{\dot{\phi}^{2}}{2} - V(\phi),
\end{equation}
where $(\rho_m, p_m)$ denote the energy density and pressure of the matter sector, while $(\rho_\phi,p_\phi)$ correspond to the scalar-field contributions. The scalar field dynamics follows from the variation of the action with respect to $\phi$, leading to the Klein-Gordon equation
\begin{equation}\label{phiminimal}
\ddot{\phi} + 3\frac{\dot{a}}{a}\dot{\phi} + \frac{dV(\phi)}{d\phi} = 0.
\end{equation}

Together, Eqs. \eqref{fe1minimal}-\eqref{phiminimal} form a closed dynamical system governing the cosmic expansion driven by a minimally coupled scalar field. For a scalar self-interaction potential of the form
\begin{equation}\label{ourpot}
V(\phi) = V_0 + \frac{\lambda}{2}\phi^2 + \frac{\epsilon}{4}\phi^4 + \frac{\eta}{2 \phi^2},
\end{equation}
the scalar-field evolution equation takes the explicit form
\begin{equation}\label{minKG}
\ddot{\phi} + 3\frac{\dot{a}}{a}\dot{\phi} + \lambda \phi + \epsilon \phi^3 - \frac{\eta}{\phi^3} = 0,
\end{equation}
and falls categorically within the Chiellini integrability framework. The structure of the potential in Eq. \eqref{ourpot} is particularly note-worthy. The quadratic and quartic self-interaction terms resemble those commonly employed in cosmological scalar-field models inspired by Higgs-like mechanisms, where spontaneous symmetry breaking and self-interaction play a central role \cite{JIMENEZ199653, Khoury_2004, Burrage_2016, PhysRevD.81.033003, Hinterbichler_2011, refId0, Sol_2016, Honardoost_2017, Peracaula_2018, PhysRevD.99.043539, PhysRevD.103.023502, Chakrabarti_2022}. At the same time, the inverse-square term is reminiscent of effective potentials appearing in axion-like theories, scalar condensates, and certain low-energy limits of quantum-corrected field theories \cite{PhysRevLett.85.1590, Moroz_2010, 14tm-ddnv, PhysRevD.93.065043}. The potential exhibits distinct asymptotic regimes, as in the large-field limit the quartic term dominates and ensures a stiff self-interaction, which stabilizes the evolution and prevents runaway behavior of the field. In contrast, in the small-field regime ($|\phi| \ll 1$), the inverse-power term becomes dominant and regulates the evolution. To write an exat solution of Eq. \eqref{minKG} using Chiellini-reduction, we define $x = t - t_0$ to find 
\begin{eqnarray}\label{eq:phi_of_t_general}
&& \phi(x) = \sqrt{\frac{1}{x^2} + \sqrt{\tfrac{p}{3}}} ~~,~~ \epsilon = -1, \\&& \label{eq:a_of_t_general}
a(t) = c_1\, \exp\big(F(x)\big)\,\big(\sqrt{3} + \sqrt{p}\,x^2\big)^{E}\,x^{1/3}, \\&&\label{H_of_t} 
H(t) = \dfrac{\textit{Num(x)}}{\big(\dfrac{\sqrt p}{\sqrt 3} + x^{-2}\big)\,x^3}, \\&&
E = \frac{1}{6} - \frac{\sqrt3\,\eta}{2\,p^{3/2}}, \\&&\nonumber
F(x) = -\frac{1}{108 p}\Big[-6(9\eta+3\lambda p-2\sqrt3\,p^{3/2}) \\&&
+ \sqrt3\sqrt p(9\eta -3\lambda p +\sqrt3 p^{3/2})\,x^2\Big]x^2, \label{defs_FE} \\&&\nonumber
\textit{Num(x)} = 1 + x^{2} + \Big(\frac{2\sqrt{3}}{3}\sqrt{p}-p\Big)x^{4} + \\&&
\Big(\frac{p}{3}-\frac{\sqrt{3}}{9}p^{3/2} - \eta\Big)x^{6}.
\end{eqnarray}
$(c_1, \eta, \lambda, p)$ are parameters that dictate the behavior of the area radius. We also introduce $E$ and $F(x)$ which isolate the slowly varying factors from $x^{1/3}$ which gives the leading small-$x$ behaviour. This is done in view of a strategy to invert the solution and write Hubble as a function of scale factor. We use a Lambert-$W$ inversion strategy which will be useful in our study of cosmological implications of the solution.

\section{Cosmological Relevance of the Chiellini framework}
\begin{figure}[t!]
\begin{center}
\includegraphics[angle=0, width=0.40\textwidth]{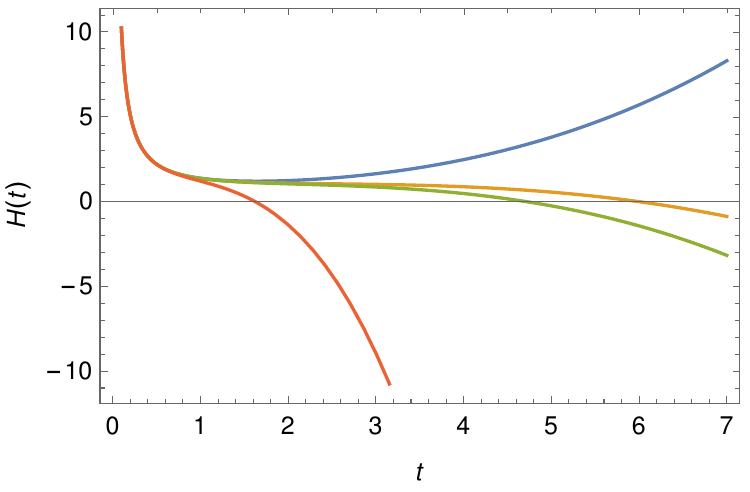}
\includegraphics[angle=0, width=0.40\textwidth]{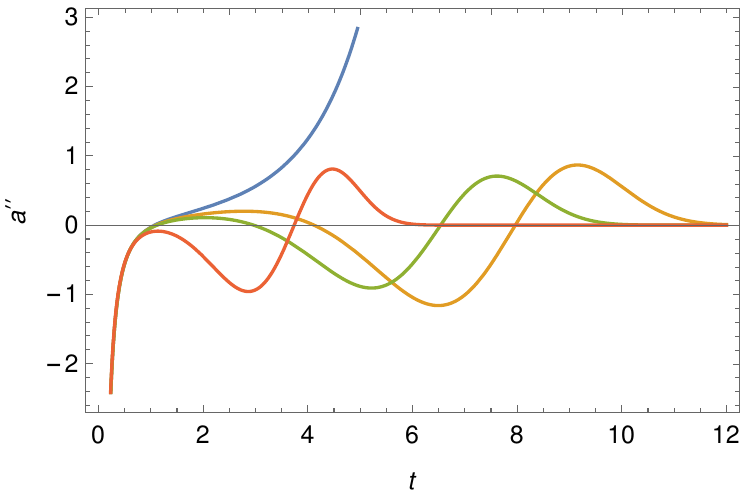}
\caption{Plot of Hubble function (top) and $\ddot{a}$ (bottom) as a function of time}
\label{Scale2}
\end{center}
\end{figure}

We first check if the solutions derived in the last section can indeed describe a cosmic expansion. We plot the corresponding Hubble function as a function of cosmic time in Fig. \ref{Scale2} (top graph). The curves in this graph show Hubble evolution for different values of $n$, while $p$ and $l$ are kept fixed. A similar qualitative beavior can also be found by varying the other parameters. There are two interesting notes to be made here : (i) depending on the choice of parameters the Hubble function can, on paper, depict an expansion with acceleration preceded by a phase of deceleration and (ii) a similar system can also be used to model the collapse a homogeneous, spherical star with the imploding branch of the solution. The collapsing model will be discussed in further details in a separate article. It is also crucial to note in passing that the exact solution can be used to describe a unified model of cosmic expansion, as in, inflation-deceleration-present acceleration. Existence of different phases of cosmic expansion is better demonstrated using an effective equation of state (EOS) $w_{\rm eff} = - 1 - \frac{2\dot{H}}{{3H}^{2}}$. A critical point defining deceleration to/from acceleration transition can be calculated from the zeros of $\ddot{a}$. We plot $\ddot{a}$ as a function of time in the bottom graph of Fig. \ref{Scale2} to show the clear evidence of multiple zero-crossing. We can evaluate the critical points of zero crossing analytically. We denote derivatives with respect to $x$ by a prime and use $\dot H = H'(x) \dot x$ and $\dot x = \tfrac{aH}{a'}$ to derive
\begin{equation}
\ddot a(x) \;=\; \frac{a(x)}{a'(x)}\,H(x)\,\frac{d}{dx}\!\big(a(x)H(x)\big).
\end{equation}
Hence the points where $\ddot a=0$ (with $a > 0$ and $a' \neq 0$) are determined by
\begin{equation}
\frac{d}{dx}\!\big(a(x)H(x)\big)=0 \quad \Longleftrightarrow\quad a'(x)\,H(x)+a(x)\,H'(x)=0.
\end{equation}

For the present model, using Eq. (\ref{eq:a_of_t_general}) the condition for inflection points of $a(t)$ can be converted into a single algebraic equation
\begin{eqnarray}\nonumber
&& \Big\lbrace F'(x)+\tfrac{2E\sqrt p\,x}{\sqrt3+\sqrt p\,x^{2}}+\tfrac{1}{3x}\Big\rbrace \mathrm{Num}(x) D(x) +  \big\lbrace \mathrm{Num}'(x)\\&&
D(x) - \mathrm{Num}(x)D'(x)\big\rbrace = 0,
\end{eqnarray}
where $D(x)=\tfrac{\sqrt p}{\sqrt3}x^3+x$. The real positive roots of the above equation marks the epochs of critical phase transitions. However, specific models of a unified cosmic expansion is not under the purview of this article and will be treated elsewhere. Our aim is to perform a statistical analysis of the Hubble function defined in Eq. (\ref{H_of_t}) and fit it with a diverse set of late-time astrophysical data. However, $H$ is primarily solved as a function of time here, not redshift. To invert the solution in Eq. (\ref{eq:a_of_t_general}) is non-trivial, which we attempt nevertheless, using a Lambert-$W$ inversion technique. 

\section{Reparametrization of the Scale Factor : a Lambert-$W$ Inversion and a rational Pade approximant}
For a a Lambert-$W$ inversion technique, we first introduce an auxiliary variable $y := x^{1/3}$ and convert Eq. \eqref{eq:a_of_t_general} into
\begin{equation}\label{eq:a_in_y}
a = c_1 e^{F(y^3)} \big(\sqrt{3} + \sqrt{p} y^6\big)^{E} y,
\end{equation}

so that the slowly varying factor can be written as an exponential of power series
\begin{eqnarray}
&& \big(\sqrt{3}+\sqrt{p} y^6\big)^{E} = (\sqrt{3})^E \exp\!\left[E\ln\!\Big(1+\tfrac{\sqrt{p}}{\sqrt{3}}y^6\Big)\right]\\&&
= A_0'\,\exp\!\big\lbrace G(y)\big\rbrace,
\end{eqnarray}
where
\begin{equation}
A_0' := (\sqrt{3})^{E} ~,~ G(y) = E\sum_{k\ge1}\frac{(-1)^{k+1}}{k} \Big(\tfrac{\sqrt{p}}{\sqrt{3}}y^6\Big)^{k}.
\end{equation}
Since $F(y^3)=\mathcal{O}(y^6)$, we can assume that the exponentials vary slowly with $y \sim x^{1/3}$. Therefore, to leading order, we absorb all constant prefactors and small exponentials into an overall constant $A_0$ and a weakly varying exponential expansion
\begin{equation}\label{eq:a_A0_By}
a = A_0\,y\,\exp\!\big(B_1 y + B_3 y^3 + B_6 y^6 + \cdots\big),
\end{equation}
where $A_0 := c_1 A_0' = c_1(\sqrt{3})^{E}$ and $B_1, B_3, B_6, \dots$ arise from expanding $F(y^3)$ and $G(y)$. In many parameter regimes, the linear coefficient $B_1$ can dominate among subleading terms. If $B_1 = 0$, the same algebraic procedure can apply for $B_3$ or the next nonvanishing coefficient replacing $B_1$. For generality, we denote the leading coefficient simply by $B$. Retaining only the dominant exponential contribution, we approximate
\begin{equation}\label{eq:a_By}
a \simeq A_0\,y\,e^{B y}.
\end{equation}
Multiplying by $B/A_0$ gives $\frac{B a}{A_0} = (B y)e^{B y}$. By definition of the Lambert function $W(z_w)$ \cite{LambertW, LOCZI2022127406}, satisfying $W e^{W} = z_w$, we find
\begin{equation}\label{eq:y_via_W}
B y = W\!\left(\frac{B a}{A_0}\right) \quad \Longrightarrow \quad y = \frac{1}{B}\,W\!\left(\frac{B a}{A_0}\right).
\end{equation}
Hence,
\begin{equation}\label{eq:x_t_via_W}
x = y^3 = \frac{1}{B^3}W\!\left(\frac{B a}{A_0}\right)^{3}, \qquad t = t_0 + x.
\end{equation}

Substituting Eq. \eqref{eq:x_t_via_W} into the parametric form for $H(t)$ yields an analytical closed form in terms of $W(z_w)$, with $z_w :=\tfrac{B a}{A_0}$. For the truncated model in Eq. \eqref{eq:a_By}, writing it as $\ln a = \ln A_0 + \ln y + B y$ and taking a direct differentiation we write
\begin{equation}
\frac{\dot a}{a} = \left(\frac{1}{y} + B \right)\dot y.
\end{equation}
Using $x = y^{3}$ and $\dot{x}=1$, we can write Hubble as
\begin{equation}\label{eq:H_of_a_W}
H = \frac{\dot a}{a} = \frac{1}{3}\big(y^{-3}+B y^{-2}\big) = \frac{B^3}{3}\left\lbrace \frac{1+W}{W^3} \right\rbrace.
\end{equation}

We define $W_1 := W(z)\big|_{a=1}$ and differentiate Eq. \eqref{eq:H_of_a_W} with respect to $a$. Using $\frac{dW}{da} = \frac{W}{a(1+W)}$, we derive the first two derivatives at $a=1$ as
\begin{eqnarray}
&& H(1) = \frac{B^3}{3} \frac{1+W_1}{W_1^3}, \label{eq:H1} \\&&
H'(1) = \frac{B^3}{3} \frac{(-2W_1-3)}{W_1^3(1+W_1)}. \label{eq:Hp1}
\end{eqnarray}

Although Eq. \eqref{eq:H_of_a_W} offers a closed analytical form, the presence of $W(z)$ can hinder direct phenomenological comparison with observational data. We derive a more practical representation by expanding $H(a)$ about $a = 1$ as $H(a) = H(1) + H'(1)(a-1) + \tfrac{1}{2}H''(1)(a-1)^2 + \tfrac{1}{6}H^{(3)}(1)(a-1)^3 + \cdots$ and capture the local analytical behaviour near the present epoch ($a \simeq 1$). We also acknowledge that truncated Taylor expansions typically converge poorly for $|a-1| \gtrsim 0.2$ and to improve global stability, we reorganize this into a rational Pade approximant \cite{Wei_2014, PhysRevD.90.043531, YU2021100772, Yu_2025}. A generic $[2/2]$ Pade form in powers of $x:= a - 1$ looks like
\begin{equation}\label{eq:pade22}
H_{\text{pade}}(a) = \frac{p_0 + p_1 x + p_2 x^2}{1 + q_1 x + q_2 x^2}.
\end{equation}
Matching this expansion to the Taylor series $H(a) = l_0+l_1x+l_2x^2+l_3x^3+l_4x^4+\cdots$, with $l_j = H^{(j)}(1)/j!$, one can find
\begin{align}\label{eq:pade_coeffs_general}
p_0 &= l_0, \qquad p_1 = -\frac{l_0 l_1 l_4 - l_0 l_2 l_3 - l_1^2 l_3 + l_1 l_2^2}{l_1 l_3 - l_2^2}, \notag\\
p_2 &= \frac{l_0 l_2 l_4 - l_0 l_3^2 - l_1^2 l_4 + 2 l_1 l_2 l_3 - l_2^3}{l_1 l_3 - l_2^2},\\
q_1 &= -\frac{l_1 l_4 - l_2 l_3}{l_1 l_3 - l_2^2}, \qquad
q_2 = \frac{l_2 l_4 - l_3^2}{l_1 l_3 - l_2^2}. \notag
\end{align}

We aim to map this reparametrized Hubble as a spatially flat $\Lambda$CDM form,
\begin{equation}\label{eq:HLambda}
H_{\Lambda}(a) = H_0\sqrt{\Omega_m a^{-3} + (1-\Omega_m)}.
\end{equation}
We normalize both of the models such that $H_{\text{pade}}(1) = H_{\Lambda}(1) = H_0$. The derivative of $H_\Lambda$ at $a = 1$ is $H_{\Lambda}'(1) = -\tfrac{3}{2}H_0\Omega_m$. By equating the slopes we find that the effective matter fraction for $H_{\text{pade}}$ can be written as
\begin{equation}\label{eq:Omega_eff}
\Omega_m^{\mathrm{eff}} = -\frac{2}{3}\frac{H_{\text{pade}}'(1)}{H_{\text{pade}}(1)} = -\frac{2}{3}\frac{s_1}{s_0}.
\end{equation}
We now define the ratio
\begin{equation}\label{eq:M_ratio}
M(u) := \frac{H_{\text{pade}}(a)}{H_{\Lambda}(a)} \bigg|_{H_0=s_0,\;\Omega_m = \Omega_m^{\mathrm{eff}}},
\end{equation}
so that $H_{\text{pade}}(a) = H_\Lambda(a)M(u)$ and $u = 1/(1+z) - 1 \equiv a - 1$. $z$ is the cosmological redshift. The function $M(u)$ brings in the pade approximant in Hubble through
\begin{equation}\label{eq:M_pade}
M(u) \approx \frac{1 + r_1 u + r_2 u^2}{1 + s_1 u + s_2 u^2}.
\end{equation}

Therefore, the inverted Hubble as a function of scale factor reads 
\begin{equation}\label{eq:H_final}
H_{\text{approx}}(a) = H_0 \sqrt{\Omega_m^{\mathrm{eff}} a^{-3} + (1-\Omega_m^{\mathrm{eff}})}\textit{P}(a),
\end{equation}
where,
\begin{eqnarray}
&& \textit{P}(a) = \frac{1 + r_1 (a-1) + r_2 (a-1)^2}{1 + s_1 (a-1) + s_2 (a-1)^2}, \\&&
H_0 = \frac{B^3}{3}\frac{1+W_1}{W_1^3} ~,~ \Omega_m^{\mathrm{eff}} = \frac{2(2W_1+3)}{3(1+W_1)^2}.
\end{eqnarray}
$W_1 = W(B/A_0)$ and constants $A_0, B$ are defined in the small-$y$ expansion. In the next section we discuss the statistical analysis of the model in Eq. (\ref{eq:H_final}) in comparison with observational data. However, for clarity we note here that the Pade approximation provides a local analytical model centered at $a=1$; broader ranges can be addressed with higher-order Pade. Also, in an observational purview, the parameters $(H_0,\Omega_m^{\mathrm{eff}},r_i,s_j')$ will be fitted directly to Hubble data. The multiplicative Pade structure will enforce correct asymptotic scaling while retaining flexibility to capture mild deviations from standard cosmology.

\begin{figure*}[t!]
\centering
\includegraphics[width=\linewidth]{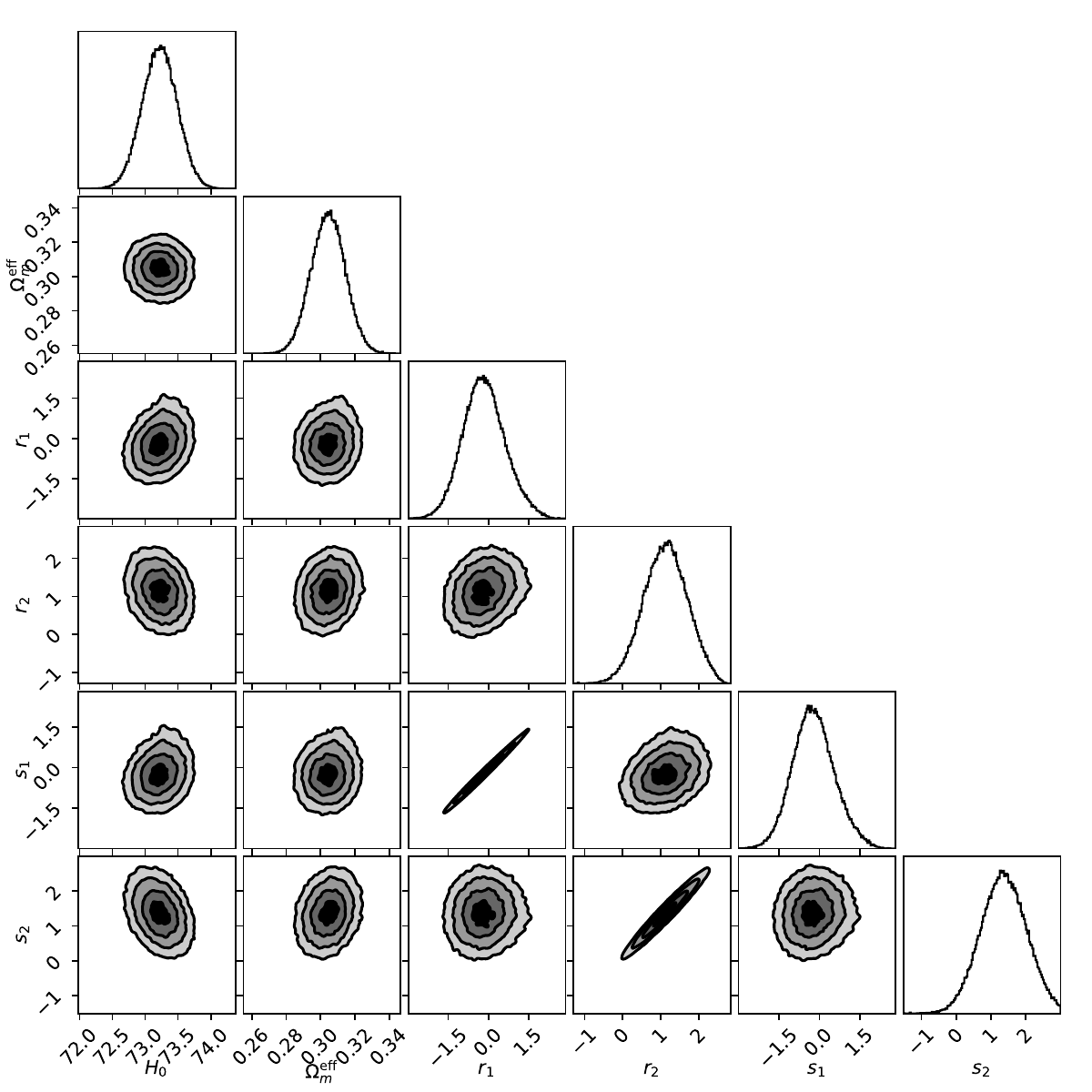}
\caption{Two-dimensional marginalized posterior distributions and one-dimensional marginalized constraints for the six-parameter cosmological model $\{H_0,\,\Omega_m^{\rm eff},\,r_1,\,r_2,\,s_1,\,s_2\}$ obtained from the combined CC + BAO + Pantheon+SH0ES + DESI + CMB dataset (Set 1). Contours correspond to the $68\%$ and $95\%$ confidence levels. The inclusion of late-time distance-ladder data leads to a preference for a higher value of $H_0$ relative to $\Lambda$CDM.
}
\label{Corner1}
\end{figure*}

\section{Comparison with Observational Data}
We perform a comprehensive Bayesian parameter estimation of the modified Hubble function, incorporating diverse cosmological probes such as cosmic chronometers (CC), baryon acoustic oscillations (BAO), type Ia supernovae (SNe) and cosmic microwave background (CMB) shift parameters. The analysis employs the affine-invariant Markov Chain Monte Carlo (MCMC) algorithm implemented in the \texttt{emcee} package \cite{ForemanMackey2013}, ensuring robust convergence across the six-dimensional parameter space $\{H_0, \Omega_m^{\rm eff}, r_1, r_2, s_1, s_2\}$.

\begin{figure}[t!]
\begin{center}
\includegraphics[angle=0, width=0.40\textwidth]{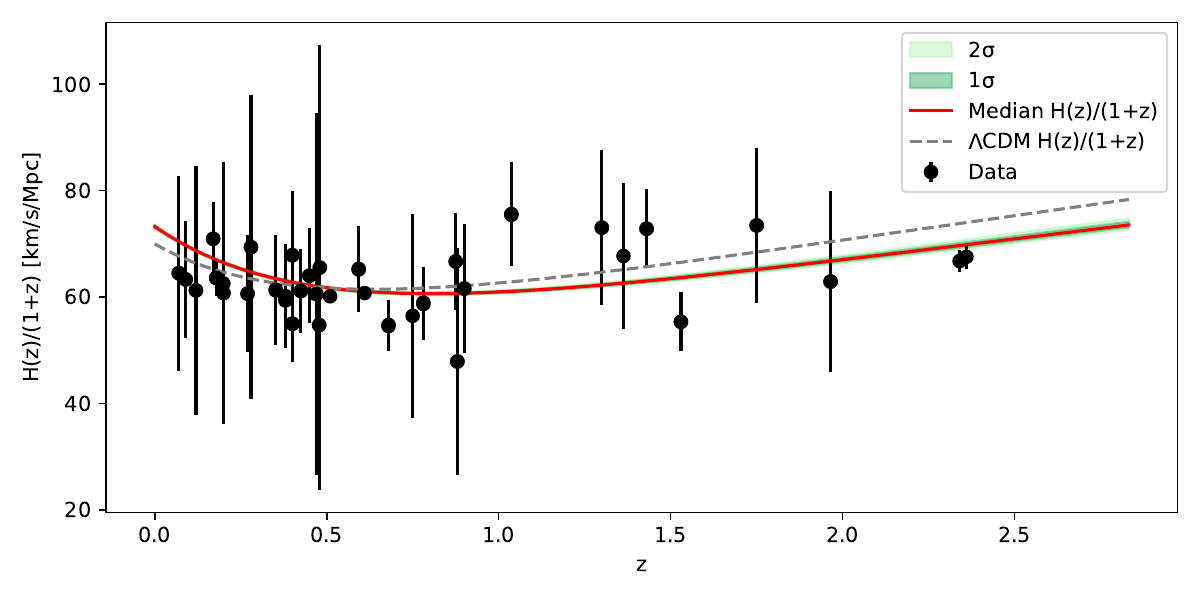}
\includegraphics[angle=0, width=0.40\textwidth]{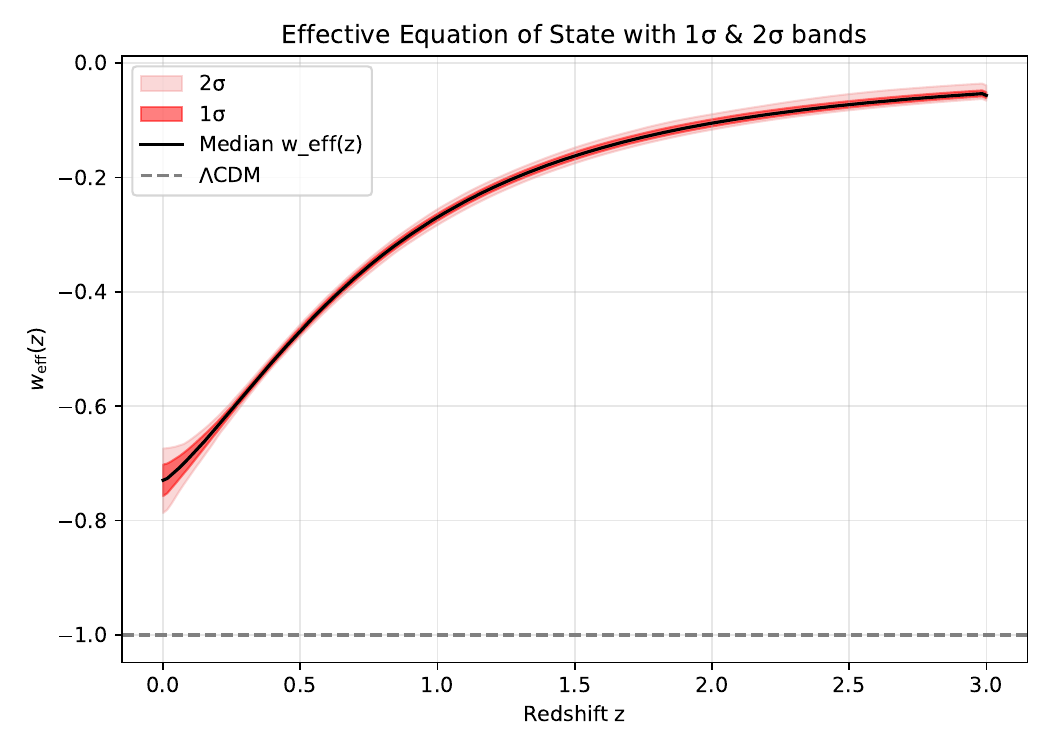}
\caption{Reconstructed cosmological evolution from the full CC+BAO+Pantheon+SH0ES+DESI+CMB data combination (Set~1). The top panel shows the rescaled Hubble parameter $H(z)/(1+z)$, and the bottom panel shows the effective equation-of-state parameter $\omega_{\rm eff}(z)$. The shaded regions denote the $1\sigma$ and $2\sigma$ confidence intervals inferred from the MCMC posterior.}
\label{Hubble1}
\end{center}
\end{figure}

The Hubble expansion rate is modeled as
\begin{equation}
H(z) = H_0\sqrt{\Omega_m^{\rm eff}(1+z)^3 + 1 - \Omega_m^{\rm eff}} 
\left\lbrace \frac{1 + r_1 u + r_2 u^2}{1 + s_1 u + s_2 u^2}\right\rbrace,
\end{equation}
where $u = 1/(1+z) - 1$ and $(r_1, r_2, s_1, s_2)$ are considered as rational polynomial deformations to the background $\Lambda$CDM expansion. The model smoothly reduces to standard $\Lambda$CDM when all deformation parameters vanish. The data sets employed in this analysis are summarized below:
\begin{itemize}
    \item \textbf{Cosmic Chronometers (CC):} Measurements of the Hubble parameter $H(z)$ obtained from the differential ages of passively evolving galaxies. This method provides a direct and model-independent probe of the expansion rate as a function of redshift. We use the compilation of $H(z)$ measurements, including redshift $z$, observed values $H(z)$, and associated uncertainties $\sigma_H$, as reported in Refs.~\cite{Jimenez2003, Moresco2016, Moresco:2020fbm}.
    
    \item \textbf{BAO:} Baryon Acoustic Oscillation distance measurements from large-scale structure surveys, including SDSS and BOSS \cite{Eisenstein2005, Anderson2014}, together with the recent DESI DR2 BAO data \cite{DESI2025}. BAO observations provide a standard ruler through the sound horizon scale $r_s$, enabling precise constraints on the expansion history. The observables are expressed in terms of ratios of the comoving angular diameter distance $D_M(z)$, the Hubble distance $D_H(z)$, and the volume-averaged distance $D_V(z)$ relative to $r_s$. We use the full covariance matrices released by the surveys to properly account for correlations among different redshift bins and distance measures.
    
    \item \textbf{Type Ia Supernovae (SNe Ia):} The Pantheon+SH0ES compilation \cite{Brout2022}, which provides standardized luminosity distance measurements over a wide redshift range. This data set plays a crucial role in constraining the late-time expansion history and the effective dark energy dynamics. Both statistical and systematic uncertainties are incorporated through the provided covariance matrices.
    
    \item \textbf{CMB:} Constraints from the Planck 2018 cosmic microwave background observations \cite{Planck2018}, implemented through the compressed likelihood based on the CMB shift parameters $(R, \ell_A, \Omega_b h^2)$. These quantities encapsulate the geometric information of the CMB and effectively constrain the physics of the early Universe and the overall matter content.
\end{itemize}

\begin{table}[t]
\centering
\caption{Best-fit parameters with $1\sigma$ uncertainties obtained from the extended six-parameter MCMC analysis.}
\begin{tabular}{|c|c|c|}
\hline
\textbf{Parameter}
& \textbf{Set 1}
& \textbf{Set 2} \\
\hline
$H_{0}$ [km\,s$^{-1}$\,Mpc$^{-1}$]
& $73.22 \pm 0.26$
& $70.26 \pm 0.57$ \\
$\Omega_{m}^{\rm (eff)}$
& $0.3049^{+0.0098}_{-0.0100}$
& $0.3083^{+0.0095}_{-0.0095}$ \\
$r_{1}$
& $-0.184^{+0.803}_{-0.747}$
& $-0.579^{+0.648}_{-0.651}$ \\
$r_{2}$
& $1.134^{+0.575}_{-0.581}$
& $0.478^{+0.603}_{-0.603}$ \\
$s_{1}$
& $-0.238^{+0.788}_{-0.732}$
& $0.522^{+0.602}_{-0.630}$ \\
$s_{2}$
& $1.363^{+0.659}_{-0.582}$
& $0.698^{+0.649}_{-0.642}$ \\
\hline
\end{tabular}
\label{tab:mcmc_results}
\end{table}

\begin{figure*}[t!]
\begin{center}
\includegraphics[width=\linewidth]{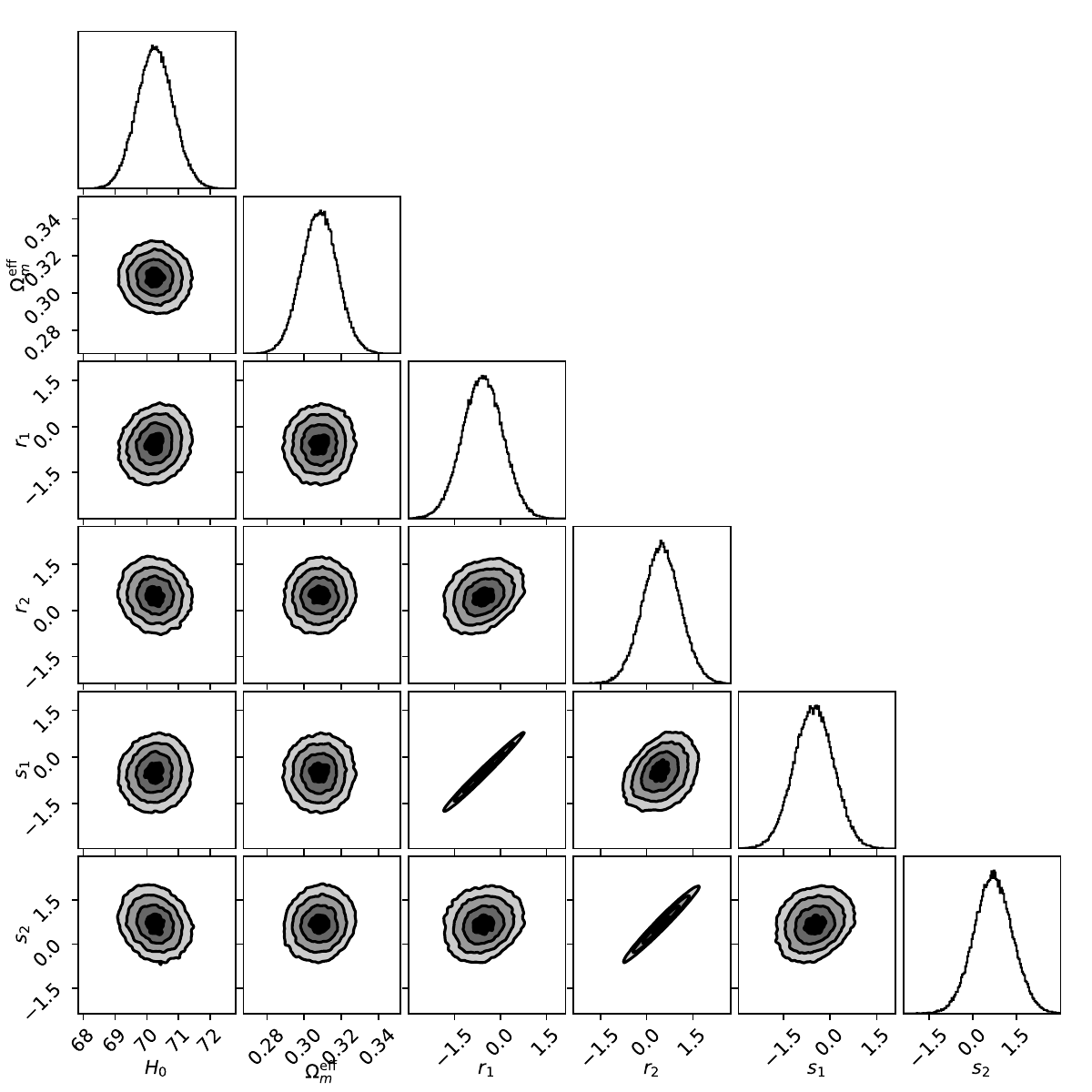}
\caption{Two-dimensional marginalized posterior distributions and one-dimensional marginalized constraints for the six-parameter cosmological model $\{H_0,\,\Omega_m^{\rm eff},\,r_1,\,r_2,\,s_1,\,s_2\}$ obtained from the CC + BAO + DESI + CMB dataset, excluding Pantheon+ (Set 2). Contours correspond to the $68\%$ and $95\%$ confidence levels. In the absence of late-time distance-ladder measurements, the inferred value of $H_0$ shifts toward lower values, consistent with Planck 2018 constraints.}
\label{Corner2}
\end{center}
\end{figure*}

\begin{figure}[t!]
\begin{center}
\includegraphics[angle=0, width=0.40\textwidth]{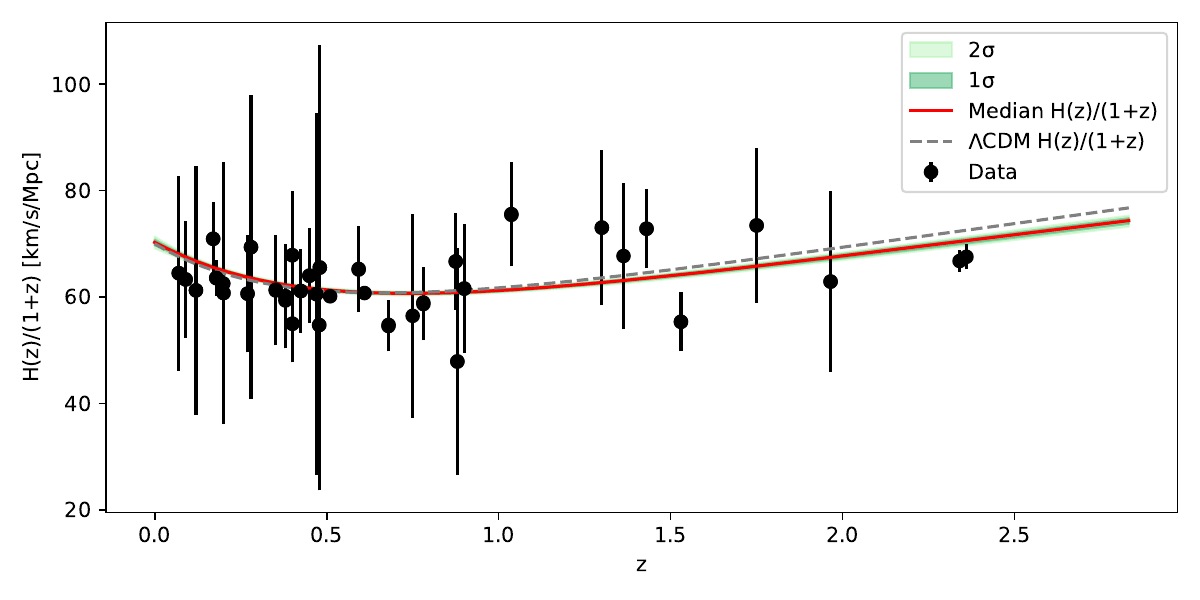}
\includegraphics[angle=0, width=0.40\textwidth]{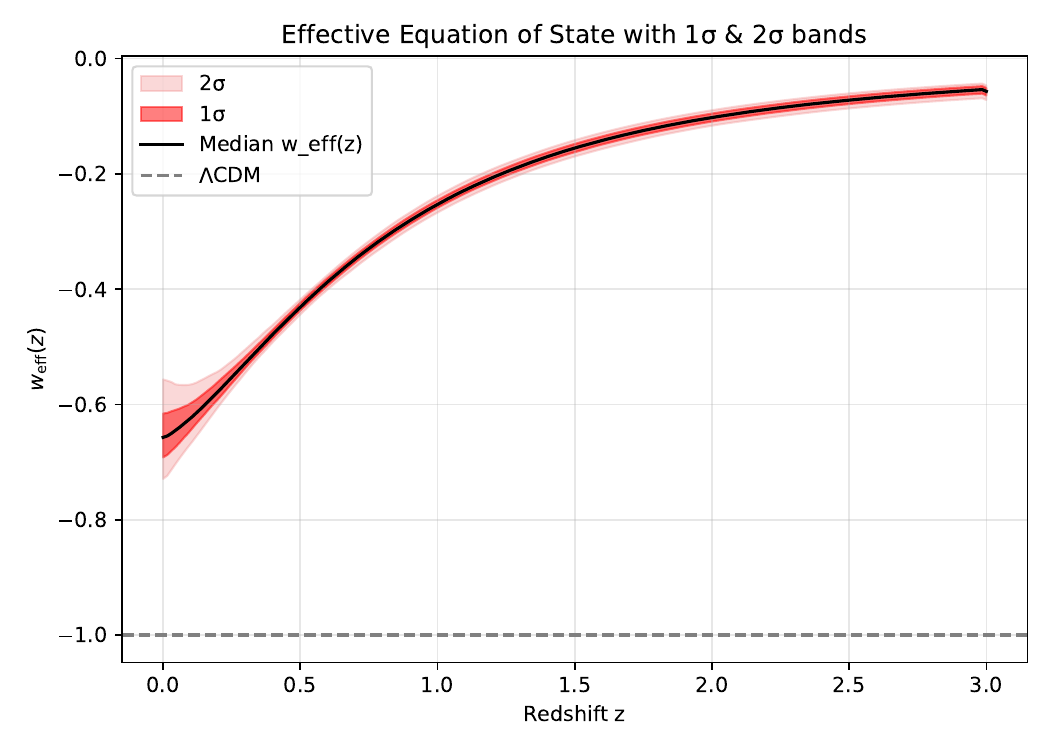}
\caption{Reconstructed cosmological evolution from the full CC+BAO+DESI+CMB data combination (Set~2). The top panel shows the rescaled Hubble parameter $H(z)/(1+z)$, and the bottom panel shows the effective equation-of-state parameter $\omega_{\rm eff}(z)$. The shaded regions denote the $1\sigma$ and $2\sigma$ confidence intervals inferred from the MCMC posterior.}
\label{Hubble2}
\end{center}
\end{figure}

We generate the two-dimensional marginalized confidence contours for the generalized Hubble using two parameter sets; in Fig. \ref{Corner1} using {\bf Set 1}: CC + BAO + Pantheon-SHOES + DESI + CMB) and in and in Fig. \ref{Corner2}, with {\bf Set 2}: CC + BAO + DESI + CMB. For the full data combination including Pantheon+SH0ES, the present-day Hubble parameter is obtained as $H_0 = 73.22 \pm 0.26~\mathrm{km\,s^{-1}\,Mpc^{-1}},$ indicating a significant alleviation of the well-known $H_0$ tension relative to the standard $\Lambda$CDM model. The corresponding effective matter density parameter is found to be $\Omega_{m}^{(\mathrm{eff})} = 0.3049^{+0.0098}_{-0.0100}$. In contrast, the analysis without Pantheon+SH0ES yields a slightly lower value, $H_0 = 70.26 \pm 0.57~\mathrm{km\,s^{-1}\,Mpc^{-1}}$, consistent with Planck 2018 constraints, thus reaffirming the sensitivity of the Hubble parameter estimation to the inclusion of late-time distance-ladder measurements. \\

The derived evolution profiles of $H(z)/(1+z)$ and the effective equation-of-state parameter $\omega_{\mathrm{eff}}(z)$ are illustrated in Figures~\ref{Hubble1} and~\ref{Hubble2} for the Set~1 and Set~2 data combinations, respectively. The evolution of $H(z)/(1+z)$ is also plotted against the observed data in Ref.~\cite{Jimenez2003}, together with the $\Lambda$CDM prediction. It can be seen that the evolution in the current $H(z)$ parametrization closely follows that of $\Lambda$CDM, although it leads to a higher value of $H_0$ in the present epoch. The evolution of $\omega_{\mathrm{eff}}$ indicates a smooth dynamical behavior, remaining close to $-1$ at low redshift and approaching $0$ at higher redshift. This suggests an appropriate transition of the Universe from matter domination into dark-energy domination. Overall, the extended six-parameter model can be considered consistent with current observational data while providing an improved fit to the late-time expansion rate. The best-fit values and corresponding $1\sigma$ uncertainties for all parameters are summarized in Table~\ref{tab:mcmc_results}. \\

\begin{figure*}
    \centering
    \includegraphics[width=\textwidth]{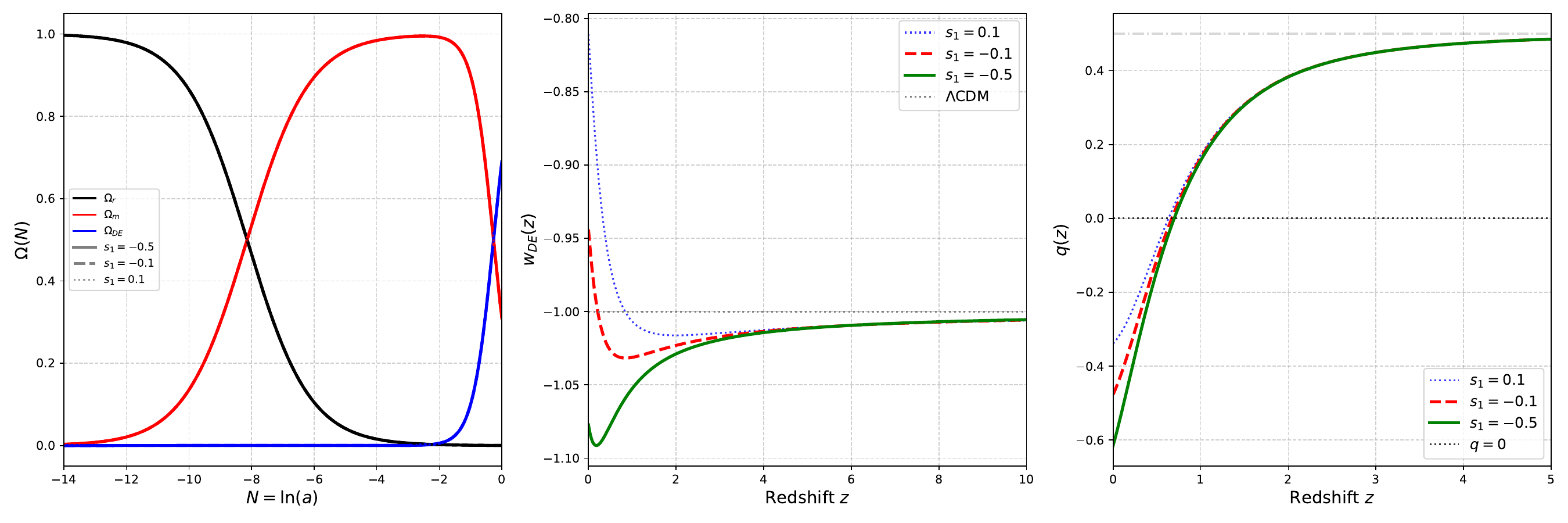}
    \caption{Summary of the reconstructed cosmological evolution in the extended six-parameter model. \textit{Left panel:} evolution of the effective matter density parameter $\Omega(N)$ as a function of the e-folding variable $N=\ln a$, showing a smooth transition from matter to dark-energy domination. \textit{Middle panel:} redshift evolution of the dark-energy equation-of-state parameter $\omega_{\rm DE}(z)$, exhibiting a low-redshift crossing of the phantom divide $\omega_{\rm DE}=-1$ within the observationally allowed region. \textit{Right panel:} evolution of the deceleration parameter $q(z)$, indicating the onset of late-time cosmic acceleration ($q<0$).
 }
    \label{fig:densities}
\end{figure*}

In Fig. \ref{fig:densities}, we present the reconstructed background evolution obtained by fixing $r_1$ to its best-fit value from the combined data set CC+BAO+Pantheon-SHOES+DESI+CM and by considering three different choices of $s_1$. The left panel shows the evolution of the effective matter density parameter $\Omega(N)$ as a function of the e-folding variable $N=\ln a$, illustrating a smooth transition from an early matter-dominated era to a dark-energy--dominated phase at late times. The middle panel displays the redshift evolution of the dark-energy equation-of-state parameter $\omega_{\rm DE}(z)$. For some parameter choices, $\omega_{\rm DE}$ crosses the phantom divide at low redshift ($z<1$), in qualitative agreement with recent results reported from DESI DR2. The right panel shows the evolution of the deceleration parameter $q(z)$, highlighting the transition from decelerated to accelerated expansion, with $q<0$ marking the onset of late-time cosmic acceleration. These plots demonstrate that the reconstructed cosmological dynamics remain smooth, regular, and free from unphysical instabilities across the entire redshift range considered, while successfully capturing both late-time acceleration and phantom-divide crossing.

It is worth emphasizing that the appearance of a phantom-divide crossing in the present model does not violate the well-known no-go theorem forbidding such a transition for a single, minimally coupled scalar field with a canonical kinetic term \cite{Vikman_2005, gcs7-c4c6}. The theorem applies to the fundamental scalar equation of state, for which crossing $w = -1$ would require a change in the sign of the kinetic energy, leading to ghost instabilities or divergent perturbations. Using the exact Chiellini-integrable solution for the scalar field,
\begin{equation}
\phi(x)=\sqrt{x^{-2}+\sqrt{p/3}}, \qquad x=t-t_0,
\end{equation}
the scalar kinetic term can be evaluated explicitly as
\begin{equation}
\dot{\phi}^2 = \frac{1}{x^4\left(1+x^2\sqrt{p/3}\right)},
\end{equation}
which is strictly positive for all finite cosmic times and decays monotonically as $x\to\infty$. As a consequence, the scalar-field EOS parameter
\begin{equation}
w_\phi = \frac{\dot{\phi}^2-2V(\phi)}{\dot{\phi}^2+2V(\phi)}
\end{equation}
never crosses the phantom divide $w_\phi = -1$, in full agreement with the standard no-go theorem for a single, minimally coupled canonical
scalar field. The scalar dynamics therefore remains ghost-free and well-defined throughout the evolution. In contrast, the effective EOS parameter governing the background expansion,
\begin{equation}
w_{\rm DE} \Omega_{DE} = -1 - \frac{2}{3}\frac{\dot H}{H^2},
\end{equation}
is controlled by the exact Hubble function obtained from the Chiellini-reduced system. Owing to the nontrivial polynomial structure of $H(t)$, the quantity $\dot H$ can change sign at finite cosmic time even though $\dot{\phi}^2 > 0$ everywhere. The phantom-divide crossing thus occurs purely at the geometric level, corresponding to a transient extremum of the Hubble rate ($\dot H=0$), rather than to any pathological behavior of the scalar sector. This behavior is naturally realized in the late-time regime, where the scalar field gradually freezes, $\dot{\phi}^2\to 0$, while subleading terms in the exact Hubble solution reorganize the expansion dynamics. The universe therefore exhibits a smooth and controlled phantom-divide crossing in $w_{\rm eff}(z)$, before asymptotically approaching a de-Sitter phase with $w_{\rm eff}\to -1$. The result provides a clear example in which an effective phantom crossing emerges from an integrable gravitational dynamics, without violating the fundamental energy conditions of the underlying scalar-field theory.

\section{Conclusion}
In this work, we have demonstrated that the Chiellini integrability condition, traditionally explored within nonlinear dissipative systems, can be systematically embedded into General Relativity through the dynamics of a minimally coupled scalar field. By reformulating the Klein-Gordon equation in a homogeneous and isotropic background, we have shown that it reduces exactly to a damped Ermakov-Painlev\'e~II equation, belonging to the broader class of generalized damped Milne-Pinney systems. This correspondence provides a rare and nontrivial example in which nonlinear scalar-field dynamics in curved spacetime can admit closed-form analytical solutions. \\

The resulting exact expressions for the scalar field, scale factor and the Hubble parameter reveal a dynamically rich expanding branch, characterized by an early decelerated phase followed by a late-time acceleration. This behavior naturally accommodates the observed accelerated expansion of the Universe without invoking ad-hoc phenomenological modifications. A key theoretical advance of this work is the reorganization of the parametric solution via a Lambert-$W$ inversion, yielding a closed-form expression for $H(a)$. When expanded around the present epoch using a rational Pad\'e approximant, this formulation produces a compact and well-behaved Hubble function consisting of a $\Lambda$CDM-like leading term multiplied by a controlled deformation. This structure enables a smooth interpolation between early- and late-time cosmic evolution while retaining analytic transparency. A comprehensive Bayesian analysis incorporating CC, BAO, DESI, CMB, and Pantheon+SH0ES data confirms the empirical viability of this framework. The extended six-parameter model yields a present-day Hubble parameter $H_0 = 73.22 \pm 0.26~\mathrm{km\,s^{-1}\,Mpc^{-1}}$, significantly alleviating the tension with late-time distance-ladder measurements, together with an effective matter density $\Omega_m^{(\mathrm{eff})} = 0.3049^{+0.0098}_{-0.0100}$. The reconstructed evolution of $H(z)$, $H(z)/(1+z)$, and the effective equation-of-state parameter $\omega_{\mathrm{eff}}(z)$ exhibits a smooth transition from deceleration to acceleration and remains close to vacuum-like behavior at low redshifts. Notably, the framework naturally allows for a smooth phantom divide crossing in the dark energy equation-of-state without pathologies, qualitatively consistent with recent indications from DESI observations, and achieved here within a single-field, integrable scalar dynamics.
\\
Overall, the Chiellini integrability condition emerges as a mathematically robust and physically consistent bridge between classical nonlinear theory and relativistic cosmology. It provides a unified analytical framework capable of describing late-time cosmic acceleration and phantom divide crossing while maintaining an excellent agreement with current observational constraints. Beyond its phenomenological success, the integrable structure uncovered here points toward a deeper organization of scalar-field cosmologies, suggesting that exact solvability may play a more central role in cosmology than usually appreciated. In a future work, this formalism can be extended into scalar-tensor theories or other modified theories of gravity and further investigate the role of integrability in governing the global structure of cosmological solutions.

\bibliographystyle{unsrt}
\bibliography{references}

@article{Milne,
  author  = {Milne, W. E.},
  title   = {The numerical determination of characteristic numbers},
  journal = {Physical Review},
  volume  = {35},
  year    = {1930},
  pages   = {863--867},
}

@article{Pinney,
  author  = {Pinney, E.},
  journal = {Proceedings of the American Mathematical Society},
  volume  = {1},
  year    = {1950},
  pages   = {681},
}

@article{PhysRevD.105.106023,
  title = {Dynamical realizations of the Lifshitz group},
  author = {Galajinsky, Anton},
  journal = {Phys. Rev. D},
  volume = {105},
  issue = {10},
  pages = {106023},
  numpages = {10},
  year = {2022},
  month = {May},
  publisher = {American Physical Society},
  doi = {10.1103/PhysRevD.105.106023},
  url = {https://link.aps.org/doi/10.1103/PhysRevD.105.106023}
}

@article{gcs7-c4c6,
  title = {Weinberg's theorem, phantom crossing, and screening},
  author = {Brax, Philippe},
  journal = {Phys. Rev. D},
  volume = {112},
  issue = {8},
  pages = {083544},
  numpages = {15},
  year = {2025},
  month = {Oct},
  publisher = {American Physical Society},
  doi = {10.1103/gcs7-c4c6},
  url = {https://link.aps.org/doi/10.1103/gcs7-c4c6}
}

@article{Vikman_2005,
   title={Can dark energy evolve to the phantom?},
   volume={71},
   ISSN={1550-2368},
   url={http://dx.doi.org/10.1103/PhysRevD.71.023515},
   DOI={10.1103/physrevd.71.023515},
   number={2},
   journal={Physical Review D},
   publisher={American Physical Society (APS)},
   author={Vikman, Alexander},
   year={2005},
   month=jan }

@article{Yu_2025,
doi = {10.3847/1538-4357/adeb81},
url = {https://doi.org/10.3847/1538-4357/adeb81},
year = {2025},
month = {jul},
publisher = {The American Astronomical Society},
volume = {988},
number = {2},
pages = {158},
author = {Yu, Bo and Liu, Wen-Hu and Yang, Xiaofeng and Zhang, Tong-Jie and Tang, Yanke},
title = {The Optimal Padé Polynomial for Reconstruction of Luminosity Distance Based on 10-fold Cross Validation},
journal = {The Astrophysical Journal}
}

@article{YU2021100772,
title = {A new analytical approximation of luminosity distance by optimal HPM-Padé technique},
journal = {Physics of the Dark Universe},
volume = {31},
pages = {100772},
year = {2021},
issn = {2212-6864},
doi = {https://doi.org/10.1016/j.dark.2021.100772},
url = {https://www.sciencedirect.com/science/article/pii/S2212686421000029},
author = {Bo Yu and Jian-Chen Zhang and Tong-Jie Zhang and Tingting Zhang}
}

@article{PhysRevD.90.043531,
  title = {Precision cosmology with Pad\'e rational approximations: Theoretical predictions versus observational limits},
  author = {Aviles, Alejandro and Bravetti, Alessandro and Capozziello, Salvatore and Luongo, Orlando},
  journal = {Phys. Rev. D},
  volume = {90},
  issue = {4},
  pages = {043531},
  numpages = {25},
  year = {2014},
  month = {Aug},
  publisher = {American Physical Society},
  doi = {10.1103/PhysRevD.90.043531},
  url = {https://link.aps.org/doi/10.1103/PhysRevD.90.043531}
}

@article{Wei_2014,
   title={Cosmological applications of Padé approximant},
   volume={2014},
   ISSN={1475-7516},
   url={http://dx.doi.org/10.1088/1475-7516/2014/01/045},
   DOI={10.1088/1475-7516/2014/01/045},
   number={01},
   journal={Journal of Cosmology and Astroparticle Physics},
   publisher={IOP Publishing},
   author={Wei, Hao and Yan, Xiao-Peng and Zhou, Ya-Nan},
   year={2014},
   month=jan, pages={045–045} }

@article{LambertW,
  title = {On the LambertW function},
  author = {Corless, R. M. et. al.},
  journal = {Advances in Computational Mathematics},
  volume = {5},
  issue = {1},
  pages = {329},
  year = {1996},
}

@article{LOCZI2022127406,
title = {Guaranteed- and high-precision evaluation of the Lambert W function},
journal = {Applied Mathematics and Computation},
volume = {433},
pages = {127406},
year = {2022},
issn = {0096-3003},
doi = {https://doi.org/10.1016/j.amc.2022.127406},
url = {https://www.sciencedirect.com/science/article/pii/S0096300322004805},
author = {Lajos Lóczi}
}

@article{Ermakov,
  author  = {Ermakov, V. P.},
  title   = {Second-order differential equations. Conditions of complete integrability},
  journal = {Univ. Izv. Kiev, Series III},
  volume  = {9},
  year    = {1880},
  pages   = {1--25},
  note    = {Reprinted/translated in Appl. Anal. Discrete Math. 2 (2008) 123--148.},
}

@misc{Haas,
  author       = {Haas, F.},
  title        = {Approximate solution for a damped Pinney equation},
  howpublished = {arXiv preprint},
  eprint       = {0712.4083},
  archivePrefix= {arXiv},
  primaryClass = {math-ph},
  year         = {2007},
}

@article{Carinena,
  author  = {Carinena, J. F. and de Lucas, J.},
  title   = {},
  journal = {International Journal of Geometric Methods in Modern Physics},
  volume  = {6},
  year    = {2009},
  pages   = {683--699},
}

@article{Rogers,
  author  = {Rogers, C.},
  title   = {A novel Ermakov--Painlev{\'e} II system, $N+1$-dimensional coupled NLS and elastodynamic reductions},
  journal = {Studies in Applied Mathematics},
  volume  = {133},
  year    = {2014},
  pages   = {214--231},
}

@article{coffman,
	author = {Coffman, C. V. and Wong, J. S. W.},
	journal = {Trans. Amer. Math. Soc.},
	volume = {167},
	pages = {399},
	year = {1972}, 
}

@article{MELLIN1994529,
title = {Solution of generalized emden-fowler equations with two symmetries},
journal = {International Journal of Non-Linear Mechanics},
volume = {29},
number = {4},
pages = {529-538},
year = {1994},
issn = {0020-7462},
doi = {https://doi.org/10.1016/0020-7462(94)90021-3},
url = {https://www.sciencedirect.com/science/article/pii/0020746294900213},
author = {Conrad M. Mellin and F.M. Mahomed and P.G.L. Leach}
}

@article{carillo,
	author = {Carillo, S. and Zullo, F.},
	journal = {Theoretical and Mathematical Physics},
	volume = {196},
	pages = {1268},
	year = {2018},
}

@article{GhoseChoudhury,
title = {Chiellini integrability condition, planar isochronous systems and Hamiltonian structures of Liénard equation},
journal = {Discrete and Continuous Dynamical Systems - B},
volume = {22},
number = {6},
pages = {2465-2478},
year = {2017},
issn = {1531-3492},
doi = {10.3934/dcdsb.2017126},
url = {https://www.aimsciences.org/article/id/ecbb1531-932e-4da0-8a10-af44114df5e0},
author = {A. Ghose Choudhury and Partha Guha}
}

@article{PhysRevD.81.033003,
  title = {Neutral Higgs-pair production at linear colliders within the general two-Higgs-doublet model: Quantum effects and triple Higgs boson self-interactions},
  author = {L\'opez-Val, David and Sol\`a, Joan},
  journal = {Phys. Rev. D},
  volume = {81},
  issue = {3},
  pages = {033003},
  numpages = {31},
  year = {2010},
  month = {Feb},
  publisher = {American Physical Society},
  doi = {10.1103/PhysRevD.81.033003},
  url = {https://link.aps.org/doi/10.1103/PhysRevD.81.033003}
}

@article{Chakrabarti_2022,
   title={Screening mechanism and late-time cosmology: Role of a Chameleon–Brans–Dicke scalar field},
   volume={514},
   ISSN={1365-2966},
   url={http://dx.doi.org/10.1093/mnras/stac1321},
   DOI={10.1093/mnras/stac1321},
   number={1},
   journal={Monthly Notices of the Royal Astronomical Society},
   publisher={Oxford University Press (OUP)},
   author={Chakrabarti, Soumya and Dutta, Koushik and Said, Jackson Levi},
   year={2022},
   month=may, pages={427–439} 
   }

@article{PhysRevD.103.023502,
  title = {Higgs chameleon},
  author = {Cai, Rong-Gen and Wang, Shao-Jiang},
  journal = {Phys. Rev. D},
  volume = {103},
  issue = {2},
  pages = {023502},
  numpages = {9},
  year = {2021},
  month = {Jan},
  publisher = {American Physical Society},
  doi = {10.1103/PhysRevD.103.023502},
  url = {https://link.aps.org/doi/10.1103/PhysRevD.103.023502}
}

@article{PhysRevD.99.043539,
  title = {Symmetron scalar fields: Modified gravity, dark matter, or both?},
  author = {Burrage, Clare and Copeland, Edmund J. and K\"ading, Christian and Millington, Peter},
  journal = {Phys. Rev. D},
  volume = {99},
  issue = {4},
  pages = {043539},
  numpages = {11},
  year = {2019},
  month = {Feb},
  publisher = {American Physical Society},
  doi = {10.1103/PhysRevD.99.043539},
  url = {https://link.aps.org/doi/10.1103/PhysRevD.99.043539}
}

@article{Peracaula_2018,
   title={Brans–Dicke gravity: From Higgs physics to (dynamical) dark energy},
   volume={27},
   ISSN={1793-6594},
   url={http://dx.doi.org/10.1142/S0218271818470296},
   DOI={10.1142/s0218271818470296},
   number={14},
   journal={International Journal of Modern Physics D},
   publisher={World Scientific Pub Co Pte Lt},
   author={Peracaula, Joan Solà},
   year={2018},
   month=oct, pages={1847029} }

@article{Sol_2016,
   title={Higgs potential from extended Brans–Dicke theory and the time-evolution of the fundamental constants},
   volume={34},
   ISSN={1361-6382},
   url={http://dx.doi.org/10.1088/1361-6382/34/2/025006},
   DOI={10.1088/1361-6382/34/2/025006},
   number={2},
   journal={Classical and Quantum Gravity},
   publisher={IOP Publishing},
   author={Solà, Joan and Karimkhani, Elahe and Khodam-Mohammadi, A},
   year={2016},
   month=dec, pages={025006} }

@article{JIMENEZ199653,
title = {Supersymmetric QCD corrections to the top quark decay of a heavy charged Higgs boson},
journal = {Physics Letters B},
volume = {389},
number = {1},
pages = {53-61},
year = {1996},
issn = {0370-2693},
doi = {https://doi.org/10.1016/S0370-2693(96)01328-7},
url = {https://www.sciencedirect.com/science/article/pii/S0370269396013287},
author = {Ricardo A. Jiménez and Joan Solà}
}

@article{Honardoost_2017,
doi = {10.1088/1475-7516/2017/11/018},
url = {https://doi.org/10.1088/1475-7516/2017/11/018},
year = {2017},
month = {nov},
publisher = {},
volume = {2017},
number = {11},
pages = {018},
author = {Honardoost, M. and Mota, D.F. and Sepangi, H.R.},
title = {Symmetron with a non-minimal kinetic term},
journal = {Journal of Cosmology and Astroparticle Physics}
}

@article{refId0,
	author = {Hagala, R. Llinares, C. and Mota, D. F.},
	title = {Cosmological simulations with disformally   coupled symmetron fields},
	DOI= "10.1051/0004-6361/201526439",
	url= "https://doi.org/10.1051/0004-6361/201526439",
	journal = {A and A},
	year = {2016},
	volume = {585},
	pages = {A37},
}

@article{PhysRevLett.85.1590,
  title = {Renormalization of the Inverse Square Potential},
  author = {Camblong, Horacio E. and Epele, Luis N. and Fanchiotti, Huner and Garc\'{\i}a Canal, Carlos A.},
  journal = {Phys. Rev. Lett.},
  volume = {85},
  issue = {8},
  pages = {1590--1593},
  numpages = {0},
  year = {2000},
  month = {Aug},
  publisher = {American Physical Society},
  doi = {10.1103/PhysRevLett.85.1590},
  url = {https://link.aps.org/doi/10.1103/PhysRevLett.85.1590}
}

@article{Moroz_2010,
   title={Nonrelativistic inverse square potential, scale anomaly, and complex extension},
   volume={325},
   ISSN={0003-4916},
   url={http://dx.doi.org/10.1016/j.aop.2009.10.002},
   DOI={10.1016/j.aop.2009.10.002},
   number={2},
   journal={Annals of Physics},
   publisher={Elsevier BV},
   author={Moroz, Sergej and Schmidt, Richard},
   year={2010},
   month=feb, pages={491–513} }

@article{PhysRevD.93.065043,
  title = {Condensates in relativistic scalar theories},
  author = {Moore, Guy D.},
  journal = {Phys. Rev. D},
  volume = {93},
  issue = {6},
  pages = {065043},
  numpages = {12},
  year = {2016},
  month = {Mar},
  publisher = {American Physical Society},
  doi = {10.1103/PhysRevD.93.065043},
  url = {https://link.aps.org/doi/10.1103/PhysRevD.93.065043}
}

@article{14tm-ddnv,
  title = {Entanglement witnesses mediated via axionlike particles},
  author = {Carmona Rufo, Pablo Guillermo and Kumar, Ayush and Sab\'{\i}n, Carlos and Mazumdar, Anupam},
  journal = {Phys. Rev. D},
  volume = {111},
  issue = {11},
  pages = {115005},
  numpages = {10},
  year = {2025},
  month = {Jun},
  publisher = {American Physical Society},
  doi = {10.1103/14tm-ddnv},
  url = {https://link.aps.org/doi/10.1103/14tm-ddnv}
}

@article{Hinterbichler_2011,
   title={Symmetron cosmology},
   volume={84},
   ISSN={1550-2368},
   url={http://dx.doi.org/10.1103/PhysRevD.84.103521},
   DOI={10.1103/physrevd.84.103521},
   number={10},
   journal={Physical Review D},
   publisher={American Physical Society (APS)},
   author={Hinterbichler, Kurt and Khoury, Justin and Levy, Aaron and Matas, Andrew},
   year={2011},
}

@article{Burrage_2016,
doi = {10.1088/1475-7516/2016/11/045},
url = {https://doi.org/10.1088/1475-7516/2016/11/045},
year = {2016},
month = {nov},
publisher = {},
volume = {2016},
number = {11},
pages = {045},
author = {Burrage, Clare and Sakstein, Jeremy},
title = {A compendium of chameleon constraints},
journal = {Journal of Cosmology and Astroparticle Physics}
}

@article{Khoury_2004,
   title={Chameleon cosmology},
   volume={69},
   ISSN={1550-2368},
   url={http://dx.doi.org/10.1103/PhysRevD.69.044026},
   DOI={10.1103/physrevd.69.044026},
   number={4},
   journal={Physical Review D},
   publisher={American Physical Society (APS)},
   author={Khoury, Justin and Weltman, Amanda},
   year={2004},
   month=feb }

@Article{sym15071416,
AUTHOR = {González Contreras, Gabriel and Yakhno, Alexander},
TITLE = {Symmetries of Systems with the Same Jacobi Multiplier},
JOURNAL = {Symmetry},
VOLUME = {15},
YEAR = {2023},
NUMBER = {7},
URL = {https://www.mdpi.com/2073-8994/15/7/1416},
ISSN = {2073-8994},
DOI = {10.3390/sym15071416}
}

@article{PhysRevD.95.024015,
  title = {Self-similar scalar field collapse},
  author = {Banerjee, Narayan and Chakrabarti, Soumya},
  journal = {Phys. Rev. D},
  volume = {95},
  issue = {2},
  pages = {024015},
  numpages = {13},
  year = {2017},
  month = {Jan},
  publisher = {American Physical Society},
  doi = {10.1103/PhysRevD.95.024015},
  url = {https://link.aps.org/doi/10.1103/PhysRevD.95.024015}
}

@article{Batic_2023,
   title={Emergence of the Gambier equation in cosmology},
   volume={38},
   ISSN={1793-6632},
   url={http://dx.doi.org/10.1142/S0217732323500311},
   DOI={10.1142/s0217732323500311},
   number={05},
   journal={Modern Physics Letters A},
   publisher={World Scientific Pub Co Pte Ltd},
   author={Batic, D. and Guha, P. and Choudhury, A. Ghose},
   year={2023},
   month=feb }

@article{Hawkins_2002,
   title={Ermakov-Pinney equation in scalar field cosmologies},
   volume={66},
   ISSN={1089-4918},
   url={http://dx.doi.org/10.1103/PhysRevD.66.023523},
   DOI={10.1103/physrevd.66.023523},
   number={2},
   journal={Physical Review D},
   publisher={American Physical Society (APS)},
   author={Hawkins, Rachael M. and Lidsey, James E.},
   year={2002},
   month=jul }

@article{GHOSECHOUDHURY2009651,
title = {On the Jacobi Last Multiplier, integrating factors and the Lagrangian formulation of differential equations of the Painlevé–Gambier classification},
journal = {Journal of Mathematical Analysis and Applications},
volume = {360},
number = {2},
pages = {651-664},
year = {2009},
issn = {0022-247X},
doi = {https://doi.org/10.1016/j.jmaa.2009.06.052},
url = {https://www.sciencedirect.com/science/article/pii/S0022247X09004880},
author = {A. Ghose Choudhury, P. Guha and B. Khanra}
}

@article{Nucci_2008,
doi = {10.1088/0031-8949/78/06/065011},
url = {https://doi.org/10.1088/0031-8949/78/06/065011},
year = {2008},
month = {dec},
publisher = {},
volume = {78},
number = {6},
pages = {065011},
author = {Nucci, M C and Leach, P G L},
title = {The Jacobi Last Multiplier and its applications in mechanics},
journal = {Physica Scripta}
}

@article{CHOWDHURY2009104,
title = {Solutions of Emden–Fowler equations by homotopy-perturbation method},
journal = {Nonlinear Analysis: Real World Applications},
volume = {10},
number = {1},
pages = {104-115},
year = {2009},
issn = {1468-1218},
doi = {https://doi.org/10.1016/j.nonrwa.2007.08.017},
url = {https://www.sciencedirect.com/science/article/pii/S1468121807001599},
author = {M.S.H. Chowdhury and I. Hashim}
}

@article{PANAYOTOUNAKOS2006634,
title = {Exact analytic solutions of the Abel, Emden–Fowler and generalized Emden–Fowler nonlinear ODEs},
journal = {Nonlinear Analysis: Real World Applications},
volume = {7},
number = {4},
pages = {634-650},
year = {2006},
issn = {1468-1218},
doi = {https://doi.org/10.1016/j.nonrwa.2005.03.025},
url = {https://www.sciencedirect.com/science/article/pii/S1468121805000611},
author = {Dimitrios E. Panayotounakos and Dimitrios C. Kravvaritis}
}

@article{Morris_2015,
doi = {10.1088/0031-8949/90/1/015202},
url = {https://doi.org/10.1088/0031-8949/90/1/015202},
year = {2014},
month = {dec},
publisher = {IOP Publishing},
volume = {90},
number = {1},
pages = {015202},
author = {Morris, R M and Leach, P G L},
title = {Symmetry reductions and solutions to the Zoomeron equation},
journal = {Physica Scripta}
}

@article{BOHMER01012010,
author = {C. G. BÖHMER and T. HARKO},
title = {NONLINEAR STABILITY ANALYSIS OF THE EMDEN–FOWLER EQUATION},
journal = {Journal of Nonlinear Mathematical Physics},
volume = {17},
number = {4},
pages = {503--516},
year = {2010},
publisher = {Taylor \& Francis},
}

@article{AC,
  author  = {Chiellini, A.},
  journal = {Bollettino dell'Unione Matematica Italiana},
  volume  = {10},
  year    = {1931},
  pages   = {301--307},
}

@article{AL,
  author  = {Lienard, A.},
  title   = {},
  journal = {Revue G{\'e}n{\'e}rale de l'{\'E}lectricit{\'e}},
  volume  = {23},
  year    = {1928},
  pages   = {901--912, 946--954},
}

@article{IB1,
  author  = {Bandic, Ivan},
  title   = {},
  journal = {Bollettino dell'Unione Matematica Italiana},
  volume  = {16},
  year    = {1961},
  pages   = {59--67},
}

@article{IB2,
  author  = {Bandic, Ivan},
  title   = {},
  journal = {Bollettino dell'Unione Matematica Italiana},
  volume  = {17},
  year    = {1962},
  pages   = {81--91},
}

@article{MH1,
  author  = {Mak, M. K. and Harko, T.},
  title   = {},
  journal = {Computers \& Mathematics with Applications},
  volume  = {43},
  year    = {2002},
  pages   = {91--94},
}

@article{MH2,
  author  = {Mak, M. K. and Chan, H. W. and Harko, T.},
  title   = {},
  journal = {Computers \& Mathematics with Applications},
  volume  = {41},
  year    = {2001},
  pages   = {1395--1401},
}

@article{MH3,
  author  = {Harko, T. and Mak, M. K.},
  title   = {},
  journal = {Computers \& Mathematics with Applications},
  volume  = {46},
  year    = {2003},
  pages   = {849--853},
}

@article{YY,
  author  = {Yurov, A. V. and Yurov, V. A.},
  title   = {},
  journal = {Journal of Mathematical Physics},
  volume  = {51},
  year    = {2010},
  pages   = {082503},
}

@misc{MH4,
  author       = {Harko, T. and Lobo, F. S. N. and Mak, M. K.},
  title        = {},
  howpublished = {arXiv preprint},
  eprint       = {1302.0836},
  archivePrefix= {arXiv},
  year         = {2013},
}

@article{RMC,
  author  = {Rosu, H. C. and Mancas, S. and Chen, P.},
  title   = {},
  journal = {Physics Letters A},
  volume  = {379},
  year    = {2015},
  pages   = {882--887},
}

@article{MR,
  author  = {Mancas, S. C. and Rosu, H. C.},
  title   = {},
  journal = {Physics Letters A},
  volume  = {377},
  year    = {2013},
  pages   = {1434},
}

@article{MR1,
  author  = {Mancas, S. C. and Rosu, H. C.},
  title   = {Integrable equations with Ermakov--Pinney nonlinearities and Chiellini damping},
  journal = {Applied Mathematics and Computation},
  volume  = {259},
  year    = {2015},
  pages   = {1--11},
}

@book{HT,
  author    = {Davis, H. T.},
  title     = {Introduction to Nonlinear Differential and Integral Equations},
  publisher = {Dover},
  address   = {New York},
  year      = {1962},
}

@book{CM,
  author    = {Bender, C. M. and Orszag, S. A.},
  title     = {Advanced Mathematical Methods for Scientists and Engineers},
  publisher = {Wiley},
  address   = {New York},
  year      = {1986},
}

@book{AD,
  author    = {Polyanin, A. D. and Zaitsev, V. F.},
  title     = {Handbook of Exact Solutions for Ordinary Differential Equations},
  publisher = {Chapman \& Hall/CRC},
  year      = {2003},
}

@article{ForemanMackey2013,
  author  = {Foreman-Mackey, Daniel and others},
  title   = {},
  journal = {Publications of the Astronomical Society of the Pacific},
  volume  = {125},
  year    = {2013},
  pages   = {306},
}

@article{Moresco2016,
  author  = {Moresco, M.},
  title   = {},
  journal = {Journal of Cosmology and Astroparticle Physics},
  volume  = {05},
  year    = {2016},
  pages   = {014},
}

@article{Jimenez2003,
  author  = {Jimenez, R. and others},
  title   = {},
  journal = {Astrophysical Journal},
  volume  = {593},
  year    = {2003},
  pages   = {622},
}

@article{Eisenstein2005,
  author  = {Eisenstein, D. J. and others},
  title   = {},
  journal = {Astrophysical Journal},
  volume  = {633},
  year    = {2005},
  pages   = {560},
}

@article{Anderson2014,
  author  = {Anderson, L. and others},
  title   = {},
  journal = {Monthly Notices of the Royal Astronomical Society},
  volume  = {441},
  year    = {2014},
  pages   = {24},
}

@article{Brout2022,
  author  = {Brout, D. and others},
  title   = {},
  journal = {Astrophysical Journal},
  volume  = {938},
  year    = {2022},
  pages   = {110},
}

@misc{DESI2025,
  author       = {{DESI Collaboration}},
  title        = {},
  howpublished = {arXiv preprint},
  eprint       = {2505.xxxxx},
  archivePrefix= {arXiv},
  year         = {2025},
}

@article{Planck2018,
  author  = {{Planck Collaboration}},
  title   = {},
  journal = {Astronomy \& Astrophysics},
  volume  = {641},
  year    = {2020},
  pages   = {A6},
}

@article{Moresco:2020fbm,
    author = "Moresco, Michele and Jimenez, Raul and Verde, Licia and Cimatti, Andrea and Pozzetti, Lucia",
    title = "{Setting the Stage for Cosmic Chronometers. II. Impact of Stellar Population Synthesis Models Systematics and Full Covariance Matrix}",
    eprint = "2003.07362",
    archivePrefix = "arXiv",
    primaryClass = "astro-ph.GA",
    doi = "10.3847/1538-4357/ab9eb0",
    journal = "Astrophys. J.",
    volume = "898",
    number = "1",
    pages = "82",
    year = "2020"
}

@article{haas2021relativistic,
  title={Relativistic Ermakov--Milne--Pinney systems and first integrals},
  author={Haas, Fernando},
  journal={Physics},
  volume={3},
  number={1},
  pages={59--70},
  year={2021},
  publisher={MDPI}
}

@article{cariglia2018cosmological,
  title={Cosmological aspects of the Eisenhart--Duval lift},
  author={Cariglia, M and Galajinsky, A and Gibbons, GW and Horvathy, PA},
  journal={The European Physical Journal C},
  volume={78},
  number={4},
  pages={314},
  year={2018},
  publisher={Springer}
}

@article{herring2007feshbach,
  title={From Feshbach-resonance managed Bose--Einstein condensates to anisotropic universes: Applications of the Ermakov--Pinney equation with time-dependent nonlinearity},
  author={Herring, G and Kevrekidis, PG and Williams, F and Christodoulakis, T and Frantzeskakis, DJ},
  journal={Physics Letters A},
  volume={367},
  number={1-2},
  pages={140--148},
  year={2007},
  publisher={Elsevier}
}

@article{esposito2019new,
  title={A new perspective on the Ermakov-Pinney and scalar wave equations},
  author={Esposito, Giampiero and Minucci, Marica},
  journal={arXiv preprint arXiv:1905.09382},
  year={2019}
}

@article{bini2020new,
  title={New solutions of the Ermakov--Pinney equation in curved space-time},
  author={Bini, Donato and Esposito, Giampiero},
  journal={General Relativity and Gravitation},
  volume={52},
  number={6},
  pages={60},
  year={2020},
  publisher={Springer}
}

@article{mukherjee2016generalized,
  title={Generalized damped Milne-Pinney equation and Chiellini method},
  author={Mukherjee, Supriya and Choudhury, A Ghose and Guha, Partha},
  journal={arXiv preprint arXiv:1603.08747},
  year={2016}
}

@article{carinena2009applications,
  title={Applications of Lie systems in dissipative Milne--Pinney equations},
  author={Cari{\~n}ena, Jos{\'e} F and De Lucas, Javier},
  journal={International Journal of Geometric Methods in Modern Physics},
  volume={6},
  number={04},
  pages={683--699},
  year={2009},
  publisher={World Scientific}
}
\appendix
\section{Mapping of MCMC Best-Fit Parameters to the Chiellini-Lambert-$W$ Solution}
\label{sec:mcmc_chiellini_map}

A central advantage of the Chiellini-integrable scalar-field solution developed in this work is that its late-time cosmological behavior admits an analytic inversion in terms of the Lambert-$W$ function. This allows a direct correspondence between the parameters inferred from cosmological data and the constants appearing in the exact solution of the underlying nonlinear system. The Lambert-$W$ parameters $(A_0, B, W_1)$ appearing in the inverted scale-factor $a \simeq A_0 y e^{B y}$, $y=x^{1/3}$, are not independent physical parameters, but composite quantities constructed from the Chiellini constants $(c_1, \eta, \lambda, p)$. As shown in Sec. $V$, the reparametrization $x \to y = x^{1/3}$ isolates the dominant algebraic behavior of the exact solution, while all remaining model dependence is absorbed into slowly varying exponential prefactors. Explicitly, the normalization and exponential slope are given by
\begin{equation}
A_0 = c_1(\sqrt{3})^{E} ~,~ E = \frac{1}{6}-\frac{\sqrt{3}\,\eta}{2p^{3/2}},
\end{equation}
while the effective coefficient $B$ arises from the leading term in the small-$y$ expansion of $F(y^3)+G(y)$ and depends on $(\lambda,p,\eta)$ only through this combination. We emphasize the fact that it is enough to form a connection of the late-time cosmological observables such as $(H_0, \Omega_m^{(\mathrm{eff})})$ with the reduced Lambert parameter defined as
\begin{equation}
W_1 = W\!\left(\frac{B}{A_0}\right),
\end{equation}
which acts as a control variable affecting qualitative evolution of the expansion history. The closed-form Hubble function is written as
\begin{equation}
H(a) = \frac{B^3}{3}\,\frac{1 + W(a)}{W^3(a)},~~ W(a) \equiv W\!\left(\frac{B a}{A_0}\right),
\end{equation}
where $W$ denotes the principal branch of the Lambert function. Evaluating at the present epoch $a = 1$, we find that
\begin{equation}
W_1 \equiv W\!\left(\frac{B}{A_0}\right).
\end{equation}

The present Hubble constant and the effective matter density parameter follow directly:
\begin{align}
H_0 &\equiv H(1) 
= \frac{B^3}{3}\,\frac{1 + W_1}{W_1^3},
\label{eq:H0_W1} \\
\Omega_m^{(\mathrm{eff})}
&= -\frac{2}{3}\frac{H'(1)}{H(1)}
= \frac{2(2W_1 + 3)}{3(1+W_1)^2}.
\label{eq:Om_W1}
\end{align}

Eqs. \eqref{eq:H0_W1}-\eqref{eq:Om_W1} establish a one-to-one mapping between the Lambert-$W$ parameter $W_1$ and the cosmological parameters $(H_0, \Omega_m^{(\mathrm{eff})}$ inferred from observations. In the numerical analysis, the late-time expansion history is reconstructed using a flexible deformation of the $\Lambda$CDM background given in Eq. \ref{eq:H_final}, where the Pad\'e or rational deformation parameters encode deviations from a pure cosmological constant. The low-redshift dynamics of the Chiellini solution is captured, to leading order, by the single quantity $W_1$, while the higher-order deformation parameters describe subleading corrections arising from the expansion of the slowly varying exponential prefactors in the exact solution. The correspondence between the MCMC-inferred parameters and the Chiellini-Lambert-$W$ quantities is summarized in Table~\ref{tab:mcmc_chiellini}.
\begin{table}[t]
\centering

\label{tab:mcmc_chiellini}
\begin{tabular}{|c|c|c|}
\hline
\textbf{\begin{tabular}{@{}c@{}}Cosmological \\ parameters\end{tabular}}
& \textbf{\begin{tabular}{@{}c@{}}Exact-solution \\ parameter\end{tabular} }
& \textbf{Relation} \\
\hline
$H_0$
& $B,\; W_1$
& $H_0 = \dfrac{B^3}{3}\,\dfrac{1+W_1}{W_1^3}$ \\
\hline
$\Omega_m^{(\mathrm{eff})}$
& $W_1$
& $\Omega_m^{(\mathrm{eff})}
   = \dfrac{2(2W_1+3)}{3(1+W_1)^2}$ \\
\hline
EOS modification
& \begin{tabular}{@{}c@{}}Higher-order terms \\ in $F(x),\,G(y)$\end{tabular}   
& \begin{tabular}{@{}c@{}}Sub leading corrections \\ to
  $a \simeq A_0\,y\,e^{B y}$\end{tabular}   \\
\hline
Phantom crossing
& Sign change of $\dot{H}$
& \begin{tabular}{@{}c@{}}Controlled by the \\ evolution of $W(a)$\end{tabular}   \\
\hline
\end{tabular}
\caption{Mapping between cosmological parameters inferred from MCMC and the parameters of the exact Chiellini--Lambert-$W$ solution.}
\end{table}

\begin{table}[t]
\centering
\caption{Illustrative correspondence between MCMC-inferred parameters and
composite quantities in the Chiellini--Lambert-$W$ framework.}
\label{tab:chiellini_numeric}
\begin{tabular}{lll}
\hline\hline
\textbf{Quantity} & \textbf{Value} & \textbf{Origin} \\
\hline
$H_0$ & $73.2\pm0.3$ km s$^{-1}$ Mpc$^{-1}$ & MCMC fit \\
$\Omega_m^{(\mathrm{eff})}$ & $0.305\pm0.010$ & MCMC fit \\[4pt]
$W_1$ & $-0.74$ & from Eq.~(84) \\
$B$ & $\sim \mathcal{O}(1)$ & from Eq.~(83) \\[4pt]
$E$ & $\sim 0.1$--$0.2$ & $(\eta,p)$ combination \\
$c_1$ & normalization & free (rescales $A_0$) \\
$\lambda,p,\eta$ & degenerate family & fixed by $B,E$ \\
\hline\hline
\end{tabular}
\end{table}
 described by,

As an example, a numerical illustration is done here by first determining $W_1$ from observationally inferred quantities, rather than postulating values for auxiliary integration constants. From the MCMC reconstruction, we obtain representative late-time values $H_0 \simeq 70~\mathrm{km\,s^{-1}\,Mpc^{-1}}$ and $\Omega_m^{(\mathrm{eff})} \simeq 0.30 $. Inverting Eq.~\eqref{eq:Om_W1} yields
\begin{equation}
\Omega_m^{(\mathrm{eff})}
= \frac{2(2W_1 + 3)}{3(1+W_1)^2}
\quad \Longrightarrow \quad
W_1 \simeq -0.74 ,
\end{equation}
where the physically admissible branch $W_1<0$ is selected to ensure accelerated expansion. With $W_1$ fixed, Eq.~\eqref{eq:H0_W1} determines the normalization constant $B$,
\begin{equation}
H_0 = \frac{B^3}{3}\frac{1+W_1}{W_1^3}
\quad \Longrightarrow \quad
B \simeq 1.05\,H_0^{1/3},
\end{equation}
where numerical factors have been absorbed for clarity. The pair $B,W_1$ therefore uniquely fixes the exact Hubble function through
\begin{equation}
H(a) = \frac{B^3}{3}\frac{1 + W\!\left(\tfrac{Ba}{A_0}\right)}{W^3\!\left(\tfrac{Ba}{A_0}\right)} .
\end{equation}

Expanding the exact solution around $a=1$, the first derivative of $H(a)$ is controlled by the evolution of the Lambert function,
\begin{equation}
\dot H \propto -\frac{1}{(1+W)^4}\left(2W+3\right),
\end{equation}
implying that the sign change of $\dot H$ and hence the phantom divide crossing occurs when $W(a)$ evolves through the critical value $W=-3/2$. For the value $W_1 \simeq -0.74$ inferred above, the exact solution predicts: accelerated expansion at the present epoch with a phantom crossing at $z \sim \mathcal{O}(0.3 \sim 0.5)$ and an asymptotic approach to $w_{\mathrm{eff}}\to -1$ at higher redshift. These features are in quantitative agreement with the MCMC-reconstructed effective equation of state and require no additional free parameters beyond those already fixed by $H_0, \Omega_m^{(\mathrm{eff})}$.  \\

This example demonstrates explicitly that the MCMC best-fit cosmology corresponds to a well-defined point in the parameter space of the exact Chiellini-integrable solution. The Lambert parameter $W_1$ acts as a control variable encoding the normalization, matter fraction, and qualitative late-time dynamics, while the additional reconstruction parameters merely capture subleading deviations around this exact background. In this sense, the MCMC reconstruction is not merely phenomenological but corresponds to a controlled truncation of an underlying exact solution of the Einstein-scalar system. This establishes a direct bridge between observational cosmology and the integrable nonlinear dynamics encoded by the Chiellini condition.

\end{document}